\documentclass[twocolumn,showpacs,preprintnumbers,amsmath,amssymb]{revtex4}

\usepackage{graphicx}
\usepackage{dcolumn}
\usepackage{bm}

\def\<{\langle}
\def\>{\rangle}

\def\a{\alpha}
\def\b{\beta}
\def\eps{\epsilon}
\def\d{\delta}

\def\k{\kappa}

\def\m{\mu}
\def\n{\nu}

\def\th{\theta}
\def\om{\omega}
\def\Om{\Omega}
\def\g{\gamma}
\def\G{\Gamma}
\def\t{\tau}

\def\del{\partial}

\def\rhodot{\mathop{\rho}^{\ \circ}}
\def\rhoddot{\mathop{\rho}^{\circ \circ}}
\def\Xdot{\mathop{X}^{\ \circ}}

\def\Rdot{\mathop{R}^{\ \circ}}
\def\Zdot{\mathop{Z}^{\ \circ}}
\def\Phidot{\mathop{\Phi}^{\ \circ}}
\def\Fdot{\mathop{F}^{\ \circ}}

\def\^{\wedge}
\def\eh{\hat{e}}
\def\thh{\hat{\th}}
\def\nh{\hat{n}}

\def\tB{\tilde{B}}
\def\tC{\tilde{C}}
\def\tD{\tilde{D}}
\def\tE{\tilde{E}}

\begin{document}

\title{ 
Dynamical Evolution of  a  Cylindrical Shell with Rotational Pressure
}

\author{Masafumi Seriu}
 \altaffiliation[Part of this work has been done at]{Institute of Cosmology, Department of Physics \& Astronomy, Tufts University,
 Medford, MA02155, USA.}
\affiliation{
Department of Physics, Faculty of Engineering, 
University of Fukui\\
Fukui 910-8507, Japan
}%
 \email{mseriu@edu00.f-edu.fukui-u.ac.jp}


\begin{abstract}
We prepare a general framework for analyzing  the dynamics  of  a cylindrical shell in the spacetime with 
cylindrical symmetry.  Based on the framework, we investigate 
a particular model of a cylindrical shell-collapse with rotational pressure, accompanying 
the radiation of  gravitational waves and massless particles.  
The model  has been introduced previously but  has been awaiting for proper analysis. 
Here the analysis is put forward: It is proved that, as far as the weak energy condition is 
satisfied outside the shell, the collapsing shell bounces back at some point irrespective 
of the initial conditions, and  escapes 
from the singularity formation. 

The behavior after the bounce depends on the sign of $p_z$, the shell pressure in the z-direction. 
When $p_z \geq 0$, the shell continues to expand without re-contraction.
On the other hand, when $p_z <0$ (i.e. it has a tension), 
the behavior after the bounce can be more complicated depending on the details of the model. However, even in this case, 
the shell never reaches  the zero-radius configuration.
 \end{abstract}
 
\pacs{04.20.Dw;04.20.-q;04.70.Bw}

\maketitle

\section{\label{sec:1}Introduction}

Investigations  on how the spacetime structures would be during and  after  
gravitational collapses  are important topics in gravitational physics and 
spacetime physics. 
Nonlinear nature of gravity is making our understanding of the phenomena 
 very difficult, both conceptually and technically. 

Steady development of numerical methods is  contributing to our deeper understanding on the phenomena. 
However, it does not mean the analytical 
investigations are becoming less meaningful. On the contrary, the more the numerical 
results we get, the more variety of analytical investigations is needed to interpret them 
 and to reach the whole understanding of the phenomena.

In this respect, we  have not yet obtained   enough variety of mathematical models 
for gravitational collapses which can be studied, mainly by analytical methods. 
Indeed, the cases of  just spherical collapses with various matter content is already 
complicated enough, and investigations are still going on~\cite{Joshi}. It can be 
inferred from this situation that constructing and analyzing other types of models may not be an easy task. 

As the next  step in this direction, cylindrical collapses have been investigated considerably. 
This class of collapsing models  are important because elongated matter distribution 
can drastically alter the destination of the gravitational collapse.  Already  in the Newtonian gravity, 
the Jeans scale plays a vital role and the elongated matter  longer than its Jeans scale  
tends to split into smaller scale objects. 
In the case of  general relativity, there is the so-called ``hoop conjecture"~\cite{hoop}; black holes are formed  
when and only when the matter of  mass $M$  is contained in a compact region whose 
circumference $C$ in any direction always satisfies $C \leq 2\pi \cdot \frac{2GM}{c^2}$. 
Therefore it is of great importance to study the fate of the collapsing elongated matter, such as 
cylindrical collapses.        

The numerical investigation of the collapse of the non-rotating dust spheroid  by Shapiro and Teukolsky~\cite{ShapTeu} 
suggested that the naked singularities could emerge when sufficiently elongated object collapses. This numerical 
result motivated further analytical investigations on the cylindrical collapses.  
Here, in view of the cosmic-censorship hypothesis~\cite{Penrose}, 
investigations on more ``natural" conditions, rather than non-rotating cases, are of significance.     
In particular, the effect of rotation and the effect of matter pressure should be studied. 
Then, an analytical model of  collapsing  cylindrical shell made of counter-rotating dust particles was investigated by 
Apostolatos and Thorne~\cite{ApThorne}. 
(Here ``counter-rotating particles" indicates the situation in which 
 the half of the particles are rotating in the opposite direction to the other half of the particles.) 
 It takes into account the rotation of particles through the rotational pressure, 
while the net angular-momentum of the shell is kept zero.   By avoiding the frame-dragging effect in this manner, 
the analysis can be relatively simple.  It  turned out that even a small amount of rotation  prevents the singularity 
formation as far as this particular model concerned.  

Another analytical model of cylindrical gravitational collapses 
was presented  and briefly studied  by Pereira and Wang~\cite{PeWa,PeWaE}. This model has  interesting features in that 
not only has it taken into account 
both the rotation and the pressure effect, but it also tries to take into account 
the radiation of gravitational waves and massless particles from the collapsing region.
Indeed, the numerical work suggests that the gravitational radiation becomes important after 
the bounce of the cylindrical shell~\cite{Piran}. 
 In spite of several interesting features, however, this model has not yet  been properly investigated so far.
 
  Since this model is what is investigated  in this paper, let us 
briefly recall their work~\cite{PeWa,PeWaE}.  Firstly  they  prepare 
general formulas in the coordinate system $\{ t, r, z, \phi  \}$ 
for matching arbitrary two cylindrical spacetimes. Here   
junction conditions~\cite{Israel} play the central role. Then they  apply the formulas to look at    
a particular model. In this model,  an interior flat  spacetime and 
an exterior cylindrical spacetime are matched together at $r=R_0(t)$.
The geometry of the exterior spacetime is  so chosen  that there is  out-going 
flux which can be interpreted as  gravitational waves and massless particles   
(see  Sec.\ref{sec:3} for the details of this model). Then a dynamical equation 
for the shell is explicitly derived (see Eq.(\ref{eq:PW-dynamical1}) in Sec.\ref{sec:4}). 
Since this non-linear equation is complicated enough, a very special class of solutions is searched for, and  
three types of special solutions are claimed to be found. They are the solutions of
(i) a collapsing shell which   bounces back  at some finite radius, followed by an  eternal expansion afterwards,
(ii) an eternally expanding shell,  and 
(iii) a collapsing shell  whose radius reaches zero, resulting in  a line-like singularity formation~\cite{PeWa,PeWaE}. 
 
The solution (iii) looks in particular interesting, since it reminds us of  the result of numerical calculations 
 by Shapiro and Teukolsky~\cite{ShapTeu}, though the spacetime structures in these two cases are not the same.
Considering several interesting features of the model, it is quite regrettable that their investigation is  
a preliminary one in nature with the  
limitation to the very special class of solutions, and moreover it turns out that the derivation of 
the special solution  faces with crucial problems, which could totally alter the main results (see Sec.\ref{sec:4} 
for the detailed account on this point).
Thus it is meaningful to analyze this model in detail.

With these situations in mind, the present paper aims mainly two goals.

 Firstly, we set up a general framework for analyzing cylindrical shell collapses 
in the same spirit as Pereira and Wang~\cite{PeWa}, but paying more attention to the mutual relations 
of three kinds of  natural frames of 
reference. They are (1) the ortho-normal frame $\{\eh_A  \}_{A=0,1,2,3}$ associated with the cylindrical spacetime geometry, 
(2) the ortho-normal frame $\{ \eh_n, \eh_\t, \eh_z, \eh_\phi \}$ 
associated with a shell $\Sigma$ in question  
(here $\eh_n$ is the normal unit 4-vector to the shell $\Sigma$, while $\eh_\t$ is the 4-vector tangent to the world-line of an 
observer located on the shell), and (3) the coordinate frame $\{ \del_t, \del_r, \del_z, \del_\phi \}$. Computations 
based on the frame (1) can be carried out systematically, while the frame (2) is more suitable for interpreting the results. 
Therefore the bridge between them would be helpful. The frame (3) has been also considered since 
it is often convenient for actual computations.  
 
The second purpose of this paper is to analyze the above-mentioned model, based on 
the framework we prepare, and investigate the dynamical behavior of the collapsing shell. 
We assume that the weak energy condition~\cite{Wald} is satisfied outside the shell. It means, roughly speaking,  that 
the collapsing matter behaves as  normal source of gravity.  
It shall be proved that, {\it the collapsing shell bounces back at some point irrespective 
of the initial conditions, and  escapes 
from the singularity formation}. 
The behavior after the bounce depends on the sign of the shell pressure in the z-direction,  $p_z$. 
When $p_z \geq 0$, the shell continues  accelerated expansion,  and its velocity 
asymptotically tends to the light velocity. 
On the other hand, when $p_z < 0$ (i.e. tension), 
the behavior after the bounce could be more complicated depending on the details of the model. 
In either case, the shell never reaches  the zero-radius configuration, contrary to the result
(the solution (iii) mentioned above) of the preliminary analysis in Refs.~\cite{PeWa} and \cite{PeWaE}.

In Sec.\ref{sec:2}, the fundamental formulas on 
the cylindrical spacetime and the 
cylindrical shell in it are derived. We in particular pay attention to the extrinsic curvature of the shell, which 
plays the  central role for studying the shell dynamics. The junction conditions, which is another essential part for 
the shell dynamics, 
are also investigated there. Based on the framework so prepared, then,  we investigate in detail a 
particular model of a collapsing cylinder in Sec.\ref{sec:3}. In particular, we prove a theorem  that 
the shell never reaches the singularity in this model as far as the weak energy condition is satisfied outside 
the shell.  Section \ref{sec:4} is devoted to the detailed comparison between the present 
analysis and the one reported in Refs.~\cite{PeWa} and \cite{PeWaE}, pinpointing why the latter had been led to 
the virtual singular solution.
Section \ref{sec:5} is devoted for summary.


\section{\label{sec:2}Geometry for a moving cylindrical shell in  
a cylindrical spacetime}


\subsection{\label{sec:2-1}The metric and frame-vectors}
\subsubsection{The metric} 
 We consider a spacetime with cylindrical symmetry defined by a line-element, 
\begin{eqnarray}
ds^2&=&-T(t,r)^2 dt^2 + R(t,r)^2 dr^2 \nonumber \\
    && \qquad \qquad + Z(t,r)^2 dz^2 + \Phi(t,r)^2 d\phi^2 \ \ .
\label{eq:ds}
\end{eqnarray}
(By a suitable coordinate transformation $(t,r) \mapsto (\tilde{t},\tilde{r})$, 
one can actually assume $T=R$, but we here keep the form Eq.(\ref{eq:ds}).)

Looking at  Eq.(\ref{eq:ds}), the theory of differential forms is 
efficiently applied~\cite{Nakahara}, by introducing natural 1-forms, 
\begin{eqnarray}
\thh^0:&=&T(t,r)\ dt, \ \ \thh^1:=R(t,r)\ dr, \ \ \thh^2:=Z(t,r)\ dz, \nonumber \\
\thh^3:&=&\Phi(t,r)\  d\phi\ \ .
\label{eq:cartan}
\end{eqnarray}
The local ortho-normal frame $\{ \eh_A \}_{A=0,1,2,3}$ is then defined as the dual to the 1-forms 
$\{\thh^A \}_{A=0,1,2,3}$. 

Fundamental geometrical quantities, such as curvatures,  w.r.t. (with respect to) 
the frame $\{ \eh_A \}_{A=0,1,2,3}$ 
are derived and collected in Appendix \ref{app:A} 
for the  convenience of later use and future applications.

Now let $\Sigma$ be  a {\it timelike} 3-surface  with cylindrical spatial symmetry 
embedded in a cylindrical spacetime described by 
the metric Eq.(\ref{eq:ds}). Due to its symmetry, the surface $\Sigma$ is characterized  by  
a single equation $r=\rho(t)$.  Let $\t$ be a proper time of an observer 
 located on the surface with $dz=d\th=0$. Then, the metric 
 Eq.(\ref{eq:ds}) yields a relation between $d\t$ and $dt$, 
\begin{equation}
d\t^2=T^2\ (1-\frac{R^2}{T^2}\  \dot{\rho}^2)\  dt^2 \ \ .
\label{eq:dt-dtau}
\end{equation}
We can use $(\t,\ z, \ \phi)$ as a coordinate system on $\Sigma$, which 
defines ortho-normal vectors tangent to $\Sigma$,
\begin{equation}
\eh_\t:=\del_\t = \frac{d t}{d \t}  \del_t + 
    \frac{d \rho}{d \t} \del_r =(X,\ \rhodot,\ 0,\ 0)_{(trz\phi)}\ \ ,
\end{equation}
and
\begin{equation}
\eh_z:=\frac{1}{Z}\del_z \ \ , \eh_\phi:=\frac{1}{\Phi}\del_\phi \ \ .
\end{equation}
Here `` ${}^\circ $ " indicates the derivative w.r.t. $\t$ and 
$X:=dt/d\t$. 
 
A normal unit vector $\nh$  of the surface $\Sigma$, 
which may also be written as $\eh_n$ or $\del_n$, and its dual one-form 
are given by, respectively,  
\begin{eqnarray}
\eh_n&=&\nh=\del_n =\left(\frac{R}{T} \rhodot,\ \frac{T}{R} X,\ 0,\ 0 
                  \right)_{(trz\phi)}\ \ \ \ , \nonumber \\
\nh_\m &=& XTR ( -\rhodot/X,\ 1,\ 0,\ 0  )_{(trz\phi)}\ \ . 
\label{eq:n-dual}                  
\end{eqnarray}
(The suffix $(trz\phi)$ indicates the frame employed.) 
We note that the set of vectors $\{ \eh_n, \eh_\t, \eh_z, \eh_\phi\}$ 
forms the local  ortho-normal frame located on $\Sigma$. 

\subsubsection{Summary of notations}

At this stage, let us summarize notations employed throughout this paper.

\begin{enumerate}
\item 
Capital Latin indices $A, \ B,\  C,\  \cdots$ are reserved for the indices w.r.t. 
the local ortho-normal frame $\{ \eh_A \}_{A=0,1,2,3}$, and runs from $0$ to $3$. 
Einstein's summation convention is adopted throughout the paper. 
\item
The first few Greek letters   
$\a$, $\b$, $\g$, $\cdots$ are reserved  for the indices w.r.t. 
the  ortho-normal frame $\{\eh_n, \eh_\t, \eh_z,\eh_\phi \}$, and   take the values $n$, $\t$, 
$z$ or $\phi$;   
The later Greek letters $\m, \n, \cdots$ are used for   
the indices w.r.t.  the coordinate bases $\{\del_t,\ \del_r,\ \del_z,\ \del_\phi \}$.
\item
Since several frames appear in the paper, we indicate  the frame explicitly 
when necessary. For instance,   $(p,q,r,s)_{(trz\phi)}:=p\: \del_t + q\: \del_r + r\: \del_z + 
s\: \del_\phi$ for a vector and 
$(p,q,r,s)_{(trz\phi)}:=p\: dt + q\: dr + r\: dz + 
s\: d\phi$ for a 1-form. 
\item
The symbols $\dot{F}$, $F'$ and  $\Fdot$  denote 
the partial derivatives of a function $F$ w.r.t. $t$, $r$  and $\t$, respectively. 
In the same manner,  $\del_n F $ indicate the directional derivative 
of $F$ w.r.t. the vector $\nh$. 
\end{enumerate} 

\subsubsection{Frequently used formulas}

We here summarize essential formulas frequently used below.
We note the relations,  
\begin{equation}
X:=\frac{d t}{d \t}= \frac{1}{ T \sqrt{1 - \frac{R^2}{T^2}\  \dot{\rho}^2 }}\ \ , \ \ 
\rhodot := \frac{d\rho}{d\t}= \dot{\rho}\  \frac{dt}{d\t} =X \dot{\rho}\ \ .
\label{eq:rhodot}
\end{equation}
Therefore it follows 
\begin{equation}
X^2 T^2 = 1+ R^2 {\rhodot\: }^2 \ \ .
\end{equation}
For any function $F(t,r)$ defined in the neighborhood of $\Sigma$, it follows 
\begin{eqnarray}
\Fdot &=& X \dot{F} + \rhodot F' = X(\dot{F} + \dot{\rho} F')  \ \ , \nonumber \\
\del_n  F &=& T^{-1}R\rhodot +  XTR^{-1}  F' \nonumber \\
          &=& XT^{-1}R (\dot{\rho} \dot{F} + T^2 R^{-2} F')  \ \ . 
\label{eq:F-formula1}
\end{eqnarray}
Conversely, we also have
\begin{eqnarray}
{\dot F} &=& X T^2 \Fdot -TR\rhodot \del_n F  \ \ ,  \nonumber \\
F' &=&  -R^2\rhodot \Fdot + XTR\  \del_n F  \ \ . 
\label{eq:F-formula2}
\end{eqnarray}
Each derivative in Eqs.(\ref{eq:F-formula1}) and (\ref{eq:F-formula2}) 
is understood to be estimated on $\Sigma$.

\subsection{
\label{sec:2-2}
Extrinsic curvature of the  surface $\Sigma$}
\subsubsection{In the ortho-normal  frame $\{ \:  \eh_\a \: \}_{\a=n,\t,z,\phi}$}

We now compute the extrinsic curvature of $\Sigma$, which later plays the essential role
in analyzing the shell dynamics. 
The extrinsic curvature is handled mainly in the {\it ortho-normal} frame 
 $\{\eh_\t, \eh_z, \eh_\phi \}$ in this paper. To remind us of it, 
 we write $K_{e_\t e_\t}$ etc. rather than $K_{\t\t}$ etc.

There is a standard formula which relates $K_{e_\a e_\b}$ with 
the Christoffel symbol $\G$ and the normal vector $\nh$, 
\begin{eqnarray}
&&K_{e_\a e_\b}= -\nh_\g \G^\g_{\a\b} =: -\G^n_{\a\b} \nonumber \\
       && \ \ \  = -{E^A}_\a {E^B}_\b {E^n}_C \G^C_{AB}
                -{E^\m}_\a {E^n}_B \partial_\m {E^B}_\b\ ,
\label{eq:Kgeneral}                
\end{eqnarray}
where in the last line, Eq.(\ref{eq:Gamma-rel}) in Appendix \ref{app:B} has been used to 
express  the Christoffel symbol $\G^\g_{\a\b}$  in terms of  $\G^C_{AB}$  since the 
latter, given in Eq.(\ref{eq:Gamma}),  is much easier to compute.

Let us first compute $K_{e_\t e_\t}$,  
\begin{equation}
K_{e_\t e_\t}= -{E^A}_\t {E^B}_\t {E^n}_C \G^C_{AB}
                -{E^\m}_\t {E^n}_B \partial_\m {E^B}_\t\ \ . 
\label{eq:Ktt}
\end{equation}
\begin{widetext}
The final result is 
\begin{eqnarray}
K_{e_\t e_\t} &=& -\frac{R}{XT} \rhoddot -\frac{R}{2XT}\rhodot\  (\ln R^2)^\circ 
                     + \frac{1}{2}(T^2X^2 -1) \  \del_n (\ln R^2) \nonumber \\
         &&\qquad  + \frac{X^2T^2}{2} (X^2T^2 - 1) \ \del_n \ln (T^2/R^2) 
                  - \frac{XTR}{2}\rhodot\  (X^2T^2 - 1) \ (\ln (T^2/R^2))^\circ \ \ .
\label{eq:Kttfinal}
\end{eqnarray}
\end{widetext}
We note that $K_{e_\t e_\t}$ contains the second time-derivative of $\rho$ as is seen in Eq.(\ref{eq:Kttfinal}), so that 
$K_{e_\t e_\t}$ plays the most important role in  the shell dynamics.  In view of its essential significance,  
the detailed derivation of Eq.(\ref{eq:Kttfinal}) is shown in Appendix \ref{app:C}. 
It is by far easier to compute $K_{e_z e_z}$ and $K_{e_\phi e_\phi}$. 
Combining Eq.(\ref{eq:Kgeneral}) with Eqs.(\ref{eq:E1}) and (\ref{eq:E3}), we
get  
\begin{eqnarray}
K_{e_z e_z} &=&  \del_n \ln Z \ \ . 
\label{eq:Kzzfinal}\\
K_{e_\phi e_\phi} &=&  \del_n \ln \Phi \ \ .
\label{eq:Kppfinal}            
\end{eqnarray}

\subsubsection{In the coordinate frame $\{\:  \del_\m \:  \}_{\m=t,r,z,\phi}$}

It is reasonable to compute every quantity  in the ortho-normal frame $\{ \eh_\a  \}_{\a=n,\t,z,\phi}$
as has been done just above.
However, in order to  compare the present results  with the results in the preceding work Ref.\cite{PeWa,PeWaE} 
(see Sec.\ref{sec:4}), it is also needed 
to represent main quantities   in terms of the coordinate frame $\{ \del_\m  \}_{\m=t,r,z,\phi}$. 

\begin{widetext}
After some computations (see Appendix \ref{app:C} for more details), we finally reach the result,
\begin{equation}
K_{e_\t e_\t} =  - X^3 T R\  \ddot{\rho} 
                + (X^2 T^2 -1)\frac{XT}{2R} 
                    \left(  (\ln T^2)' + \frac{R^2}{T^2}\  \dot{\rho}\  (\ln R^2\dot{)\ }   
                    \right)    
          + \frac{1}{2}{X^2 TR }\  \dot{\rho}\  ( \ln (T^2 / R^2)  )^\circ  \ \ . 
\label{eq:Kttfinal2}
\end{equation}
\end{widetext}
Here  the $\tau$-derivative in  the last term on the R.H.S. (right-hand side) has been purposefully left untouched  
  to make the formula look simpler; the term will vanish whenever $T=R$, as is the case for 
the model we analyze later.

\subsection{
\label{sec:2-3}Junction Conditions}

Discontinuity in the extrinsic curvature across a 
3-surface implies that the energy-momentum is accumulated on the shell; this is 
analogous to the relation of the discontinuity in the electric field 
across a shell with the  electric charge there. 

Geometrical discontinuity  across a 3-surface $\Sigma$ is described by 
a discontinuity-tensor induced on $\Sigma$ defined as 
\begin{equation}
\k S_{\a \b}=
\left[ K_{\a \b}  \right]  - h_{\a \b}\left[ K  \right] \ \ ,
\label{Sab}
\end{equation}
where $\left[ Q  \right]:=Q_+ - Q_-$, the difference of $Q$ across 
the surface; $\k:=8\pi G$ is the Einstein's gravitational constant. 

Then the discontinuity is  governed by the junction conditions~\cite{Israel},
\begin{equation}
{\rm (I)} \ \ \ {ds_{-}^2}_{|\Sigma} = {ds_{+}^2}_{|\Sigma}\ \  \ \ \ \ 
{\rm (II)} \ \ \ D_\b {S_\a}^\b = \left[ T_{\m \n} {E^\m}_\a \nh^\n  \right]\ \ .
\label{eq:junction}
\end{equation}
Here $T_{\m \n}$ is the energy-momentum tensor in the neighborhood of 
$\Sigma$; $\nh^\m$ and ${E^\m}_\a$ are given in Eqs. (\ref{eq:n-dual}) and 
 (\ref{eq:E1}), respectively, if one refers to  the $(t, r, z, \phi)$-frame.

Let us now prepare   more explicit expression for Eq.(\ref{eq:junction}). 
The metric induced on the surface $\Sigma$ is 
\begin{eqnarray*}
ds_\Sigma^2 &=& -d\t^2 +Z^2 dz^2 + \Phi^2 d\phi^2  \\
             &=& -\thh^\t \otimes \thh^\t + \thh^z \otimes \thh^z 
            + \thh^\phi \otimes \thh^\phi \ \ , 
\end{eqnarray*}
where $\thh^\t = d\t$, $\thh^z = Z dz$, and $\thh^\phi= \Phi d\phi$  
form a set of ortho-normal 1-forms on $\Sigma$ dual to the bases
$\{\eh_\a  \}_{\a=\t, z,\phi}$. 
Thus the discontinuity-tensor in this frame becomes
\begin{eqnarray}
S_{e_\t e_\t} &=& 
        - \frac{1}{\k} \left(\left[ K_{e_z e_z}  \right] 
                            + \left[ K_{e_\phi e_\phi}  \right] 
                       \right) =: \eps\ \ , \nonumber  \\
S_{e_z e_z} &=& 
        - \frac{1}{\k} \left(\left[ K_{e_\t e_\t}  \right] 
                            - \left[ K_{e_\phi e_\phi}  \right] 
                       \right) =: p_z \ \ , \nonumber  \\
S_{e_\phi e_\phi} &=& 
        - \frac{1}{\k} \left(\left[ K_{e_\t e_\t}  \right] 
                            - \left[ K_{e_z e_z}  \right] 
                       \right) =: p_\phi \ \ , 
\label{eq:e-p-p}                        
\end{eqnarray}
where $\eps$, $p_z$ and $p_\phi$ are interpreted as 
the energy density of the shell, the pressure in the $z$-direction and 
pressure in the $\phi$-direction, respectively. 
It is easy to get 
\begin{eqnarray*}
&& \G^{e_\t}_{e_z e_z}= (\ln Z)^\circ \ , \ 
\G^{e_\t}_{e_\phi e_\phi}= (\ln \Phi)^\circ \  ,  \
\G^{e_z}_{e_z e_\t}= (\ln Z)^\circ \ \ , \\
&& \G^{e_\phi}_{e_\phi e_\t} = (\ln \Phi)^\circ \  , \ 
 {\rm others} = 0 \ \ . 
\end{eqnarray*}
(We note $\{ \eh_\t  \eh_z  \eh_\phi\}$ forms the non-coordinate bases and,  
in particular, $\G^{\eh_z}_{\eh_\t \eh_z}= \G^{\eh_\phi}_{\eh_\t \eh_\phi}= 0$.)
Thus
\begin{equation}
 D_\b {S^\b}_\a 
 = (-\mathop{\eps}^{\ \circ} -(\eps + p_z)(\ln Z)^\circ - 
    (\eps+p_\phi) (\ln \Phi)^\circ,\  0,\  0 ) 
\label{eq:DSab}    
\end{equation}
in the $(\thh_\t  \thh_z \thh_\phi)$-coordinate.
Then (II) in Eq.(\ref{eq:junction}) reduces to 
\begin{equation}
\mathop{\eps}^{\ \circ} +(\eps + p_z)(\ln Z)^\circ + 
    (\eps+p_\phi) (\ln \Phi)^\circ 
    = -\left[ T_{\m \n}\: {E^\m}_\t \:  \nh^\n  \right]\ \ , 
\label{eq:junc-1}
\end{equation}
with  $\nh^\n$ and ${E^\m}_\a$ being given in 
Eq.(\ref{eq:n-dual}) and Eq.(\ref{eq:E1}), respectively.

Noting that 
\begin{eqnarray*}
S_{\cdot \cdot} &=& \eps {\thh^\t} \otimes {\thh^\t}
          + p_z{\thh^z} \otimes {\thh^z}  
          + p_\phi{\thh^\phi}_a \otimes {\thh^\phi}  \\
    &=& \eps d\t \otimes d\t 
          + p_z Z(\t)^2 dz \otimes dz   
          + p_\phi \Phi(\t)^2 d\phi \otimes d\phi \ \ , \\
\end{eqnarray*}
We also get 
\begin{eqnarray*}
S_{ab} = diag (\eps,\  Z(\t)^2 p_z\ , \Phi(\t)^2 p_\phi )_{(\t, z, \phi)} \ \ . \\
\end{eqnarray*}
It is straightforward to get 
\begin{eqnarray*}
&& \G^\t_{zz}=\frac{1}{2} (Z^2)^\circ \ ,  \ 
\G^\t_{\phi \phi}=\frac{1}{2} (\Phi^2)^\circ  \ ,  \
\G^z_{\t z}=\G^z_{z \t}=  (\ln Z)^\circ \  ,  \\
&& \G^\phi_{\t \phi}= \G^\phi_{\phi \t}= (\ln \Phi)^\circ \ \ , 
 {\rm others} = 0 \ \ ,  
\end{eqnarray*}
and
\begin{eqnarray*}
D_b {S^b}_a 
  = (-\mathop{\eps}^{\ \circ} -(\eps + p_z)(\ln Z)^\circ 
    -(\eps+p_\phi) (\ln \Phi)^\circ,\  0,\  0 ) 
\end{eqnarray*}
in the $(\t, z, \phi)$-coordinate, 
which is  the same expression appears as in Eq.(\ref{eq:DSab}).
Therefore, (II) in Eq.(\ref{eq:junction}) reduces to Eq.(\ref{eq:junc-1}) 
in the $(\t, z, \phi)$-frame also.

\section{
\label{sec:3}
Analysis of a collapsing shell model with rotational pressure and 
gravitational radiation}

Having prepared all the necessary formulas in the preceding sections, 
let us now reanalyze the cylindrical collapsing-shell model considered 
by Pereira and Wang~\cite{PeWa}.  
The spacetime to be analyzed is constructed by matching 
two spacetimes ${\cal M}_{\pm}$  described by the metrics $ds_\pm$  
at the timelike surface $\Sigma$ (the suffix $-$ is for the interior geometry
 and $+$ is for the exterior geometry): 
The $ds_\pm$  are given by 
\begin{eqnarray}
ds_-^2 &=& - d{t_-}^2 + d{r_-}^2 + d{z_-}^2 + {r_-}^2 d{\phi_-}^2  ,
\label{eq:ds-}
 \\
ds_+^2 &=& e^{2\g (\xi)} (-d{t_+}^2 + d{r_+}^2) + d{z_+}^2 + {r_+}^2 d{\phi_+}^2,
\label{eq:ds+}
\end{eqnarray}
where $\g =\g (\xi)$, a function of $\xi:=t_+-r_+$ only: The surface $\Sigma$ is 
assumed to be described by $r_{\pm}=\rho_{\pm}(t)$.  The metric $ds_+$ has the form
of $ds$ in Eq.(\ref{eq:ds}) with 
\begin{eqnarray*}
T(t_+,r_+) &=& R(t_+,r_+) = e^{\g (\xi)} \  , \ \ Z(t_+,r_+) =1\  , \\
\Phi(t_+,r_+) &=& r_+\ \ . 
\end{eqnarray*}

The $C$-energy (cylindrical energy)~\cite{CEnergy, ApThorne} of this 
spacetime is given by
\begin{equation}
C = \frac{1}{8 \k }(1-e^{-2\g}) \ \ ,
\label{eq:C-energy}
\end{equation}
so that we assume $\g \geq 0$ from now on.

The interior spacetime is obviously flat, while the exterior geometry 
is given by, with the help of Eq.(\ref{eq:Ricci}) and  
noting that $\dot{\g} = - \g'$, $\ddot{\g} = \g''$,   
\begin{eqnarray*}
{\bf R}_{00}&=&{\bf R}_{11}= \frac{\g'}{r_+} e^{-2\g}\ \ , \\
{\bf R}_{01} &=& {\bf R}_{10}= \frac{\dot{\g}}{r_+} e^{-2\g} = -\frac{\g'}{r_+} e^{-2\g} \ \ , \\
 {\rm others} &=& 0\ \ . 
\end{eqnarray*}
Therefore 
${\bf R}=0$ and
${\bf G}_{00}={\bf G}_{11}=-{\bf G}_{01}=-{\bf G}_{10}=\frac{\g'}{r_+} e^{-2\g}$, or 
\begin{eqnarray*}
  {\bf G}_{\cdot \cdot} &=& \frac{\g'}{r_+} e^{-2\g} (\thh^0 \otimes \thh^0 -\thh^0 \otimes \thh^1 
              -\thh^1 \otimes \thh^0 + \thh^1 \otimes \thh^1)  \\
        &=& \frac{\g'}{r_+} (dt_+ \otimes dt_+ - dt_+ \otimes dr_+   - dr_+ \otimes dt_+ \\
        && \qquad + dr_+ \otimes dr_+)  \ \ .     
\end{eqnarray*}
Thus~\cite{PeWa} 
\begin{equation}
{\bf G}_{ab}= \frac{\g'}{\a^2  r_+ } k_a k_b\ \ , 
\label{eq:EinsteinTensor}
\end{equation}
where $k_\m = (\a,\ -\a,\ 0,\ 0)_{(t r z \phi)}$ is a null-vector.

Combined with the Einstein equation, Eq.(\ref{eq:EinsteinTensor}) 
implies  $\g' \geq 0$ when the weak energy condition for matter is imposed. 
Since we are interested in the behavior of the collapsing shell in 
more or less ``natural" conditions in view of the cosmic censorship conjecture~\cite{Penrose},  
it is reasonable to assume $\g \geq 0$ and $\g' \geq 0$ from now on. 

It may be appropriate to make some remarks on the energy condition. 
Without any condition imposed on matter, any strange spacetime  would become possible. Indeed, 
one can always claim that  arbitrary spacetime is the solution of the Einstein equation if 
the matter with  energy-momentum tensor  $T_{ab} := \frac{1}{\k} {\bf G}_{ab} $ is prepared. 
Therefore it is essential to assume some kind of energy condition  to make any meaningful arguments~\cite{Wald}.
The weak energy condition is one of such appropriate  conditions widely accepted, so we assume the condition 
here to make the definite arguments below.  Of course, there is still a room of considering the violation of 
the energy condition at the shorter scale, due to the Casimir effect for instance. Though such a possibility is 
certainly interesting to study, we do not consider it  in this paper for definiteness and only investigate 
the purely classical situations.

We now impose the junction conditions Eq.(\ref{eq:junction}) for two geometries ${\cal M}_\pm$:
\vskip .7cm
\fbox{Junction Condition (I)}
\vskip .4cm
We impose 
\begin{equation}
{{ds_-^2}_|}_{r_-=\rho_-(t_-)} = {{ds_+^2}_|}_{r_+=\rho_+(t_+)}\ \ .
\label{eq:junction(I)}
\end{equation}
When $dz_{\pm}=d\phi_{\pm}=0$, it implies 
\begin{eqnarray}
d\t^2=
\left\{
       \begin{array}{ccc} 
         dt_-^2 - dr_-^2 & = & (1-\dot{\rho}_-^{\: 2})dt_-^2  
\label{eq:Junction(I')}  \\
         e^{2\g_+} (dt_+^2 - dr_+^2) &=& e^{2\g_+} (1-\dot{\rho}_+^{\: 2})dt_+^2
       \end{array}
\right.     
\nonumber   
\end{eqnarray}
where 
$\g_+:=\g(t_+,\rho_+(t_+))$.
By plugging the relation
$\dot{\rho}_-^{\: 2} = \left( \frac{dt_+}{dt_-}  \right)^2 \dot{\rho}_+^{\: 2}$ into 
the above relation and by solving the latter w.r.t. 
$\frac{dt_+}{dt_-}$,  we get an important formula 
\begin{equation}
\frac{dt_+}{dt_-} 
 =  \left\{ (1 - e^{2\g_+} ) \dot{\rho}_+^{\: 2} + e^{2\g_+}   \right\}^{-1/2} =:\Delta^{-1} \ \ .
\label{eq:dTdt}
\end{equation}
Taking the $t$-derivative of the above relation, one finds 
\begin{equation}
\frac{d^2t_+}{dt_-^2}  
=- \frac{1}{\Delta^4}
  \left\{
     \dot{\rho}_+ (1- e^{2\g_+}) \ddot{\rho}_+ 
       - e^{2\g_+} \g_+' (1- \dot{\rho}_+^{\:2}) 
  \right\}\ \ .
\label{eq:dt2dt2}  
\end{equation}
Noting the relation,
\begin{equation}
\frac{d}{dt_+} \Delta
= \frac{1}{\Delta} \dot{\rho}_+ \ddot{\rho}_+ 
    - \frac{e^{2\g_+}}{\Delta}(\dot{\rho}_+ \ddot{\rho}_+ 
      +\g_+' (1- \dot{\rho}_+^{\: 2} ) ) \ \ ,   
 \label{eq:rhominusdot}
\end{equation}
one also finds
\begin{eqnarray}
\ddot{\rho}_-(t_-) 
&=& \Delta^{-1} \frac{d}{dt_+} \left( \Delta^{-1} \dot{\rho}_+  \right) \nonumber \\
&=& \frac{e^{2\g_+}}{\Delta^4}
    \left\{
        \ddot{\rho}_+ 
       + \g_+' ( 1- \dot{\rho}_+^{\: 2} ) \dot{\rho}_+ 
    \right\} \ \ .
\label{eq:rhominus2dot}    
\end{eqnarray}

\vskip .7cm
\fbox{Junction Condition (II)}
\vskip .4cm
Let us find the expressions for $K_{e_\a e_b}$ (Eqs.(\ref{eq:Kzzfinal}), (\ref{eq:Kppfinal}) and (\ref{eq:Kttfinal2})) 
in this model, and then derive the formulas for $\epsilon$, $p_z$ and $p_\phi$ (Eq.(\ref{eq:e-p-p})). 

First of all, the factor $X$ becomes
\[
X= \frac{e^{-\g}} {  \sqrt{1-\dot{\rho}^{\: 2}} }\ \ ,
\]
implying that 
\[
\dot{\rho}^{\: 2} < 1 \ \ .
\]
Then, noting the relation 
$X_- = \Delta X_+$, Eq.(\ref{eq:Kttfinal2}) with $T^2=R^2=e^{2\g}$ yields
\begin{eqnarray*}
{{K_{e_\t e_\t}}_|}_+ &=& 
          -{X_+}^3 e^{2\g_+} \ddot{\rho}_+    +  \g_+' {X_+}^3 e^{2\g_+}  \dot{\rho}_+^2(1-\dot{\rho}_+) \ \ , \\
{{K_{e_\t e_\t}}_|}_- &=& -{X_-}^3  \ddot{\rho}_-  \\
                      &=& - \frac{{X_+}^3}{\Delta} e^{2\g_+}  \ddot{\rho}_+
                              -  \frac{\g_+'}{\Delta} {X_+}^3 e^{2\g_+}   (1-{\dot{\rho}_+}^2) \dot{\rho}_+ \ \ , 
\end{eqnarray*}
where Eq.(\ref{eq:rhominus2dot}) has been used in the last line. 
Noting also 
\begin{eqnarray*}
K_{e_z e_z} &=& 0 \ \ , \\
K_{e_\phi e_\phi} &=& \del_n \ln r = \frac{X}{r} \ \ , 
\end{eqnarray*}
\begin{widetext}
one finds
\begin{eqnarray*}
\left[ K_{e_\t e_\t}  \right]&=& -\frac{e^{-\g_+}}{\Delta (1-{\dot{\rho}_+}^2)^{3/2} } 
                                  \left\{
                                      (\Delta - 1)\ddot{\rho}_+ 
                                     - \g_+'  \dot{\rho}_+ (1 - \dot{\rho}_+)[(\Delta + 1) \dot{\rho}_+ + 1 ] 
                                  \right\} \ \ , \\  
\left[ K_{e_z e_z}  \right] &=& 0 \ \ , \\ 
\left[ K_{e_\phi e_\phi}  \right] &=& - e^{-\g_+} \frac{\Delta - 1}{ \rho_+ \sqrt{1-{\dot{\rho}_+}^2}} \ \ .
\end{eqnarray*}
Equation (\ref{eq:e-p-p}) then yields
\begin{eqnarray}
\k \eps &=& e^{-\g_+} \frac{\Delta - 1}{ \rho_+ \sqrt{1-{\dot{\rho}_+}^2}}\ \ ,
\label{eq:eps-result}    \\
\k p_z  &=&  \frac{e^{-\g_+}}{\Delta (1-{\dot{\rho}_+}^2)^{3/2} }  
   \left\{
                                      (\Delta - 1)\ddot{\rho}_+ 
                                     -\g_+'  \dot{\rho}_+ (1 - \dot{\rho}_+)[(\Delta + 1) \dot{\rho}_+ + 1 ] 
                                     -\Delta ( \Delta -1) \frac{1-{\dot{\rho}_+}^2}{\rho_+}
                                     \right\} \ \ ,  
\label{eq:pz-result} \\  
\k p_\phi &=&    \frac{e^{-\g_+}}{\Delta (1-{\dot{\rho}_+}^2)^{3/2} } 
                     \left\{
                                      (\Delta - 1)\ddot{\rho}_+ 
                                    -\g_+'  \dot{\rho}_+ (1 - \dot{\rho}_+)[(\Delta + 1) \dot{\rho}_+ + 1 ] 
                                     \right\} \ \ ,
\label{eq:pphi-result}                                     
\end{eqnarray}
with $\Delta = \left[{\dot{\rho}_+}^2 + e^{2\g_+} (1 - {\dot{\rho}_+}^2)\right]^{1/2} $.
\end{widetext}

Noting the relation $\Delta^2-1 = (1-\dot{\rho}^2)(e^{2\g} -1)$, 
we see 
\begin{equation}
\g_+
\begin{array}{c}
 >  \\
 =   \\
 <
\end{array}
0  
\Longleftrightarrow
\Delta 
\begin{array}{c}
 >   \\
 =   \\
 <
\end{array}
1 
\Longleftrightarrow
\eps 
\begin{array}{c}
 >   \\
 =   \\
 <
\end{array}
0\ \ .
\label{eq:g-Delta-eps}
\end{equation}
When $\g \equiv 0$, $\Delta =1$ and $\eps = p_z =p_\phi =0$, as it should be.

Setting 
\[
P_0:= \k p_z\ \ ,
\]
Eq.(\ref{eq:pz-result}) yields the dynamical equation of the shell, 
\begin{widetext}
\begin{equation}
\ddot{\rho}_+ = \Delta \frac{1-{\dot{\rho}_+}^2}{\rho_+} 
                  +\g_+' \frac{\dot{\rho}_+}{\Delta - 1}  (1 - \dot{\rho}_+) [(\Delta+1) \dot{\rho}_+ + 1] 
                 +\frac{\Delta}{\Delta-1}(1-\dot{\rho_+}^2 )^{3/2} e^{\g_+} P_0  \ \ .
\label{eq:result1}
\end{equation}
\end{widetext}
To make the statements concise, 
we indicate the first, the second and the third terms on the R.H.S. of Eq.(\ref{eq:result1}) 
by the symbols, $[[1]]$, $[[2]]$ and $[[3]]$, respectively.
 
The term $[[3]]$  indicates that when 
the shell has a positive pressure in the $z$-direction ($P_0 >0$), 
it causes more acceleration in the expanding direction, while the tension 
of the shell ($P_0 <0$) works to reduce  the  acceleration. However the  term $[[3]]$   
does not change the whole dynamics so much: The term could become important 
only when $\Delta \sim 1$,  i.e. when $\dot{\rho}_+ \sim \pm 1$.  Setting $\dot{\rho}_+ = \pm (1 -\d)$ ($\d >0$), 
the term behaves as $\frac{e^\g_+}{e^{\g_+} -1}  \d^{1/2}$, so that the contribution of the term is not
 significant. 
 Indeed, we shall show below a theorem that under the dynamical equation 
Eq.(\ref{eq:result1}), the shell never reaches the singular configuration 
$\rho_+ =0$ irrespective of the initial conditions nor the value of $P_0$.

First of all, we  note the following obvious fact:
\begin{description}
\item{\bf Lemma 1} \\
{\it
Consider the case $P_0 \geq 0$. 
Once the shell satisfies the condition $\dot{\rho}_+ \geq 0$, 
it expands forever afterwards.} 
\subitem{\it Proof:}\\ 
Noting that $\Delta >1$ and that the  terms $[[2]]$ and  $[[3]]$ 
are non-negative  when $\dot{\rho}_+ \geq 0$, it follows that 
$\ddot{\rho}_+ > \frac{1-{\dot{\rho}_+}^2}{\rho_+} > 0$, so that  
the claim follows.
\fbox{}
\end{description}

In particular, when $|\dot{\rho}_+| \ll 1 $,  Eq.(\ref{eq:result1}) becomes 
$
\ddot{\rho}_+ \sim e^{\g_+}/\rho_+  
\label{eq:result1b}
$, so that the effective potential behaves like $C_{eff} \sim -e^{\g_+}\log \rho_+ + \  const.$  
in the region where $\g_+$ does not change so drastically.  Compared with 
the dust case~\cite{ApThorne}, where $C_{eff} \sim a/\rho_+^2  + b\ln \rho_+  +\  const.$ ($a$, $b$ are 
constants), the tendency of expansion of the present model is obvious.

Next, let us investigate the contracting shell.  
To avoid any confusion, let us set $V:=-\dot{\rho}_+$, so that 
$0<  V < 1$. 
Then Eq.(\ref{eq:result1}) reads 
\begin{widetext}
\begin{equation}
\ddot{\rho}_+ = \Delta \frac{1-V^2}{\rho_+} 
                  + \g_+' \frac{V}{\Delta - 1}  (1 + V) [(\Delta+1)V - 1]
                  + \frac{\Delta}{\Delta -1}(1-V^2)^{3/2} e^{\g_+} P_0   \ \ ,
\label{eq:result1b}
\end{equation}
with $\Delta = \left[ V^2 + e^{2\g_+}(1-V^2)   \right] ^{1/2}$. 
\end{widetext}

The present model has been constructed by giving metrics Eqs. (\ref{eq:ds-}) and  (\ref{eq:ds+}), 
and there is no particular physical image regarding  the matter content forming the shell.  
However, one can infer that the  term $[[1]]$ comes from 
the angular momentum effect since it is a decreasing function of $\rho_+$. On the other hand,  
the term $[[2]]$  may be identified with  what corresponds to Newtonian gravity under suitable conditions,
since it is proportional to $\g_+'$.  

Let us pay attention to the last factor in the term $[[2]]$, 
 $f(V):=(\Delta+1)V - 1$.  The function $f(V)$ is obviously monotonic, increasing, 
 continuous function of $V$, 
 with  $f(0)=-1$ and $f(1)= 1$, so that $f(V)$ has only one zero in $(0,1)$. 
 
When the weak energy condition is satisfied, and the shell is not moving so fast ($V \ll 1$), then  
$f(V) < 0 $,  
so that the  term $[[2]]$  yields an attractive force, which 
is consistent with the attractive nature of Newtonian gravity. 
A curious feature arises, however, when $V \sim 1$. Then $f(V) >0  $, and  the same term 
now yields a repulsive force. 
Due to this characteristic behavior of $f(V)$, the shell bounces back at some point, and   
never collapses to singularities as we shall see soon. 

Before proving this statement, let us first  get rough understanding on  what happens when $\rho_+$ becomes small. 
The term $[[1]]$ is the repulsive force,  produced by  an  infinite effective potential  
barrier near  $\rho_+ =0$ (angular momentum effect). In case the shell would have ever reach $\rho_+ =0$, it should 
have been this repulsive effect becomes negligible compared to other terms, i.e.  when $\dot{\rho_+}$ approaches $-1$ ($V \rightarrow 1$)  more rapidly  than $\rho_+$ approaches to zero. 
 However, in this case, the term $[[2]]$
 causes a very strong repulsive effect since $f(V) > 0 $  and $\Delta \sim 1$, while the term $[[3]]$ becomes 
 negligible. 
This behavior can be analyzed in more detail as follows. 
Let us set $V = 1 - \d$ ($\d >0$). Then $\Delta = 1 + (e^{\g_+} -1) \d +O(\d ^2) $, and one finds 
$[[1]] \sim \frac{\d}{\rho_+}$, $[[2]]\sim \d^{-1}$, and $[[3]]\sim \d^{1/2}$, so that 
\[
\frac{\rho_+ \ddot{\rho}_+}{\dot{\rho}_+^2} \sim  
 \frac{2\g_+'}{e^{\g}-1 } \frac{\rho_+}{\d} \ \ .
\]
Therefore,  when $\d /\rho_+ \rightarrow 0 $, then  
the term $[[2]]$ causes a very strong repulsive effect. 

 In summary, the shell never reaches $\rho_+ =0$ due to the strong repulsive force caused by 
 the term $[[1]]$  (when $V$ is not so large), or by the term $[[2]]$  (when $V \sim 1$).

To make the statements below as concise as possible, let us first  define the ``core-region".
Consider the phase space $\G:=\{ (\rho_+, \dot{\rho}_+) |  \rho_+ \geq 0,  -1 \leq \dot{\rho}_+ \leq 1 \}$. 
It is understood that suitable  topology is endowed on $\G$.\footnote{The most natural topology 
is the one  induced from the  canonical structure read by the Einstein-Hilbert action 
constructed from the metric Eq.(\ref{eq:ds+}). The detailed topological property is not 
very important in the argument below, so that it suffices to 
 imagine $\G$  as $(radial-coordinate) \times {\bf R}$.} 
At each time $t$, the region $\cal C$ is defined such as 
 (i) $[[2]]+[[3]] >  0$ and $\dot{\rho}_+ < 0 $ on $\cal C$ and  
 (ii) $\cal C$ is  connected,  and contains  a neighborhood of $(\rho_+=0, \dot{\rho}_+=-1)$.
 In other words, $\cal C$ consists of phase points that are ``close enough" to the singularity-emerging point 
 $(\rho_+=0, \dot{\rho}_+=-1)$.        
    
\begin{description}
\item{\bf{Definition}} \\
We call the above-mentioned region $\cal C$ in $\G$  
{\it the core-region} (at  $t$), for brevity.     

We also say that the shell ``enters (or, leaves) the core-region  at $t=t_*$" 
if at  $t=t_* -0$ the shell's phase point  is outside (or, inside) the core-region, and 
at $t=t_*$, the phase point of the shell is on the edge of the core-region; 
($[[2]]+[[3]] =  0$ and $\dot{\rho}_+ < 0$). 
\end{description}   

It is clear (from the consideration of the order of magnitude of the terms $[[1]]$, $[[2]]$ and $[[3]]$ when $V \sim 1$)  that, if the shell 
would have ever reached the singularity,  the phase point of the shell 
should have reached the point $(\rho_+=0, \dot{\rho}_+=-1)$  {\it through the core-region}.

Now  we can show  
\begin{description}
\item{\bf{Lemma 2}} \\
{\it
Under the dynamical equation Eq.(\ref{eq:result1}), 
once the shell enters the  core-region at $t=t_*$, 
it bounces back without  reaching zero-radius.  
Indeed,  $\rho_+(t)$ is bounded from below as
\[ 
\rho_+(t) > \rho_+(t_*)\sqrt{1-{\dot{\rho}_+(t_*)}^2 } \ \ .
\]
}
\subitem{\it Proof:}\\ 
Noting that $\ddot{\rho}_+ >  \frac{1-{\dot{\rho}_+}^2}{\rho_+}$ when the shell is in the core-region, 
it suffices to show the  claim holds even for 
\[ 
\ddot{\rho} = \frac{1-{\dot{\rho}}^2}{\rho} \ \ ,
\]
with arbitrary initial conditions $\rho(0)=a (>0)$ and $\dot{\rho}(0) = -b $ (where $a>0$ and $1> b>0$).
However, its solution is given by 
$\rho(t) = ((t-ab)^2+  + a^2 (1-b^2) )^{1/2}$, so that $\rho_+(t) > \rho(t)\geq  a\sqrt{1-b^2} >0$. 
Thus the claim follows. 
\fbox{}
\end{description}

\begin{description}
\item{\bf{Theorem 1}} \\
{\it
Under the dynamical equation (with or without $P_0$) Eq.(\ref{eq:result1}), 
the shell never reaches $\rho_+=0$ irrespective of its initial conditions.
}
\subitem{\it Proof:}\\ 
It suffices only to consider the contracting phase. Then if the shell 
would have ever reached  $\rho_+ =0$,  the phase point of the shell 
should have reached the point $(\rho_+=0, \dot{\rho}_+=-1)$   through the core-region.  
However,  {\it Lemma 2} indicates that this never happens. 
\fbox{}
\end{description}

Thus {\it the line-like singularity never forms in this model irrespective of the value $P_0$, as far as 
the weak energy condition is satisfied outside the shell}.

The behavior in the expanding phase (and/or after the bounce) varies depending on the value $P_0$. 
When $P_0 \geq 0 $, $\ddot{\rho}_+ > 0$ in the expanding phase, so that the 
shell continues to be accelerated and expand,  and its velocity asymptotically approaches to the light 
velocity.  
On the other hand, the case $P_0 < 0$, the  term $[[3]]$ is negative, thus it might be 
possible that $\ddot{\rho}$ turns to negative when the shell velocity is not very large. 
Therefore, the oscillating behavior can arise, which might be interesting to investigate further. 
(However, even in this case, the shell never collapses to zero-radius due to 
{\it Theorem 1}.)
Once the shell velocity becomes sufficiently large in the expanding phase, however, 
the shell continues to be accelerated and expands forever  just as the case of $P_0 \geq 0 $.   
This is because,   
when the shell velocity becomes close to 1,  by  setting  $\dot{\rho}_+ = 1 - \d $ ($\d >0 $, $\d \ll 1$), 
the  term $[[2]]$ behaves as $\sim \d^{-1}$ and become dominant, while 
the  terms $[[1]]$ and  $[[2]]$ behave as $\sim \d$.

Now let us illustrate what has been proved rigorously above by some numerical investigations. 
We set $P_0=0$ for simplicity. 

 Figure 1 shows a typical behavior of the  contracting shell (thinner curve)  along with $p_\phi$ (thicker curve).
  As an example, we have set $\g_+(r_+-t_+) = \frac{1}{10^6}(r_+ - t_+ +100)^3$. 
 Initial conditions have been set $\rho_+(0) = 0.1$ and $\dot{\rho}_+(0)= -0.999$ 
 (i.e. very close to the light velocity).   The bouncing behavior discussed above is obviously seen in the 
 figure.  
 It is also seen that the turning point  of the shell coincides with the peak of the rotational pressure.

Figure 2 shows the behavior of the same model 
with the initial conditions $\rho_+(0) = 0.1$ and $\dot{\rho}_+(0)= -0.1$ (i.e. relatively mild contraction). 
Again, evolutions of $\rho_+(t)$ (thinner curve) and $p_\phi$ (thicker curve) are indicated. 
 Due to the mild initial conditions, the shell soon bounces back.
 Just as the previous case,  the turning point  of the shell coincides with the peak of the rotational pressure.  

\begin{figure*}
  \includegraphics{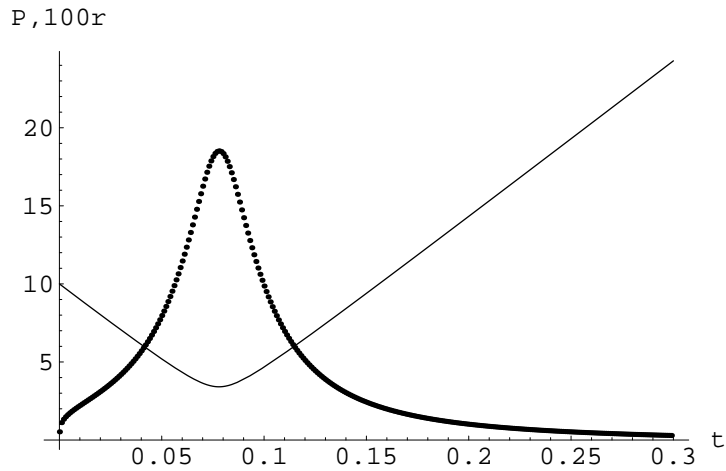}
\caption{Typical evolutions of the  shell-radius (thinner curve) and the rotational pressure $p_\phi$ (thicker curve). 
We have set $\g_+(r_+-t_+) = \frac{1}{10^6}(r_+ - t_+ +100)^3$.
Initial conditions are $\rho_+(0) = 0.1$ and $\dot{\rho}_+(0)= -0.999$. 
The vertical line indicates $100 \rho_+$ and $p_\phi$, while   the horizontal line indicates  $t_+$.}
\end{figure*}

\begin{figure*}
  \includegraphics{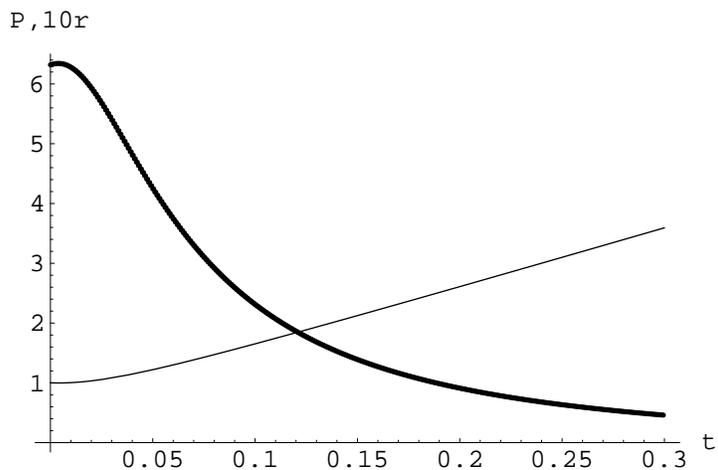}
\caption{The same model as FIG. 1 with  initial conditions, $\rho_+(0) = 0.1$ and $\dot{\rho}_+(0)= -0.1$. 
The vertical line indicates $10 \rho_+$ (for the thinner curve) and $p_\phi$ (for the thicker curve), 
while   the horizontal line indicates  $t_+$.}
\end{figure*}

Noting that $\left[ K_{e_z e_z}  \right] = 0$, Eq.(\ref{eq:e-p-p}) with  
$p_z=0$ implies $\eps = p_\phi$, so that 
\begin{eqnarray}
\k \eps &=& \k p_\phi - \k p_z =\frac{e^{-\g_+} (\Delta - 1)}{ \rho_+ (1-{\dot{\rho}_+}^2)^{1/2} } - P_0
\nonumber  \\
\k p_\phi &=&  \frac{e^{-\g_+} (\Delta - 1)}{ \rho_+ (1-{\dot{\rho}_+}^2)^{1/2} }  \ \ . 
\label{eq:result2}
\end{eqnarray}
This relation provides another  confirmation of the validity of  the condition $\g > 0$ (see Eq.(\ref{eq:g-Delta-eps})). 

Finally let us note that the relation Eq.(\ref{eq:result2}) reveals a  curious character of the model. 
We first consider the case $P_0 =0$.
In the case of the counter-rotating dust cylinder, there is a relation 
$\frac{p_\phi}{\eps} =v^2$, where $v$ is the velocity of the dust particle relative to an observer 
``standing still" on the shell~\cite{ApThorne}. Comparing this with Eq.(\ref{eq:result2}) ($P_0=0$), 
the present model   corresponds to the case of  $v=1$, so that the present model  could be  interpreted  as    
a cylinder shell  consisting of the  counter-rotating massless particles, though the two models are not exactly the same. 
The situation in which   massless particles are confined to a compact region  
 arises naturally in  superconducting objects, so that the present model should have 
some relevance.  
On the other hand, when $p_z <0$, it behaves more or less like a normal matter and it is consistent with 
the attractive nature of the term $[[3]]$. On the other hand, when $p_z >0$, 
Eq.(\ref{eq:result2}) rather looks like  a cosmic string.  However, the cylinder tends to 
expand in the present case, contrary to the case of cosmic strings.   It may not 
 appropriate at present to pursue such comparisons too further.

\section{
\label{sec:4}
Comparison with the results of the preceding paper}
 
In view of our results above,  we need to investigate whether the singularity-forming solution claimed in 
Ref.\cite{PeWa} and its correction Ref.\cite{PeWaE} is a valid one. It turns out that 
the formal solution in question is not a relevant one and should be discarded. 

By comparing one by one our formulas with the corresponding ones shown in Refs.\cite{PeWa} and \cite{PeWaE}, 
it turns out that some formulas in Ref.\cite{PeWa} along with its correction Ref.\cite{PeWaE} still 
need slight  modifications (see below and Appendix \ref{app:D} for more details).  However, 
we first derive the above conclusion faithfully based on the formulas in Refs.\cite{PeWaE} and \cite{PeWa}. Then 
 we also show that the same conclusion follows based on  our  formulas, too, even with the $P_0$-term.
 
Firstly their functions $f$, $g$, $h$, $l$, $R_0(T)$, ${R_0}'(T)$,  ${R_0}''(T)$ and $b(T-R)$ correspond to ours as 
\begin{eqnarray}
&& f:=T^2\  ,  \  g:=R^2\  ,   \ 
h:= Z^2 \  ,  \ l:= \Phi^2 \  , \nonumber  \\
&& R_0(T):=\rho(t)\  , \  {R_0}'(T):=\dot{\rho}(t)\  ,  \nonumber \\
&& {R_0}''(T):=\ddot{\rho}(t)  \ , \ b(T-R)=-2\g(t-r) \ \ .
\label{eq:notationmatch}
\end{eqnarray}
In particular, let us note that the sign of the function $b$ is opposite to our function $\g$.  Thus we should understand that  
$b \leq 0$ and $b' \leq 0$. They  correspond to the physical conditions 
$\g \geq  0$ and $\g' \geq 0$ (see arguments after Eq.(\ref{eq:C-energy}) and Eq.(\ref{eq:EinsteinTensor})).

The detailed comparison is shown in Appendix \ref{app:D}. Here it suffices to pay attention only to 
their dynamical equation shown in Ref.\cite{PeWaE} (which is the correction of  the original Eq.(30) of Ref.\cite{PeWa}):
\begin{equation}
{R_0}''=(1-{{R_0}'}^2) 
      \left\{
      \frac{\Delta}{R_0} + \frac{b'(\xi_0) (1- {R_0}') }{2 ({\Delta -1})}
       ({R_0}'- \Delta) \right\}  \ \ , 
\label{eq:PW-dynamical1}
\end{equation}
where $R_0$ and its derivatives correspond to our  $\rho_+(t)$ and its $t$-derivatives 
(see Eq.(\ref{eq:notationmatch})), 
 and $b(\xi_0)$ corresponds to our $-2{\g_+}$; $\Delta := \left[{{R_0}'}^2 + e^{-b}(1-{{R_0}'}^2)   \right] ^{1/2}$. 

 If the terms  are rearranged to look as similar to 
Eq.(\ref{eq:result1}) (with $P_0=0$)  as possible,  their formula Eq.(\ref{eq:PW-dynamical1}) reads 
\begin{eqnarray}
 && {R_0}''  = \Delta \frac{1-{{R_0}'}^2}{{R_0}}  \nonumber \\ 
          && \ \ \   +\frac{(-b'(\xi_0))}{2(\Delta -1)}(1-{{R_0}'}^2) (1- {R_0}') (\Delta - {R_0}')\ .
\label{eq:PW-dynamical2}
\end{eqnarray}
(Compared  with Eq.(\ref{eq:result1}) ($P_0=0$), 
the factors $(1-{{R_0}'}^2)$ and $(\Delta - {R_0}')$ of the second term should be 
replaced by ${R_0}'$ and $((\Delta+1){R_0}' +1 ) $, respectively.) 

Based on Eq.(\ref{eq:PW-dynamical1}), 
Ref. \cite{PeWaE} (just as Ref. \cite{PeWa})  tries to search for  special solutions for $R_0$ satisfying 
\begin{eqnarray}
&& {R_0}''=  \b \  \frac{1-{{R_0}'}^2}{{R_0}'}   \ \ , \label{eq:PW-special1} \\
&&  \b := {R_0}' \left[
                   \frac{\Delta}{R_0} + \frac{(-b') (1- {R_0}') }{2 (\Delta -1) }   ( \Delta -  {R_0}') 
                \right] ,
\label{eq:PW-special2}                
\end{eqnarray}
where $\b$ is assumed to be an arbitrary {\it constant}.

  As is  claimed in Ref.\cite{PeWa}, 
Eq.(\ref{eq:PW-special1}) itself is formally satisfied by                   
\begin{eqnarray}
 {R_0}'(T) =  - (1-e^{-2\b (T-T_0)})^{1/2} \ \ , 
\label{eq:PWsol0}
\end{eqnarray}
or integrating once more, 
\begin{eqnarray}
 R_0(T) &=& \{R_* - (T-T_0)\}  \nonumber \\
 && + \frac{1}{\b}  \Big\{ (1-e^{-2\b (T-T_0)})^{1/2} \nonumber \\
      && \qquad       -\ln [ 1+ (1- e^{-2\b (T-T_0)})^{1/2} ]  \Big\}\ \ .
\label{eq:PWsol}
\end{eqnarray}
Here two integral constants $T_0$ and $R_*$ have appeared corresponding to 
the fact that the differential equation Eq.(\ref{eq:PW-special1}) is of the 2nd order. 
The constant $T_0$ indicates the time when ${R_0}'$ vanishes, while $R_*$ (along with $T_0$) determines 
the time $T^*$  when ${R_0}$ vanishes ($T^* \leq T_0$). These constants $T_0$ and $R_*$ are 
(indirectly) determined by the initial conditions for solving Eq.(\ref{eq:PW-special1}). 

Equation (\ref{eq:PWsol}) would imply  a line-like  singularity formation,  reaching $R_0 =0$ at some finite time $T^*$, 
so that its implication would be significant. 
We show below, however,  that  {\it Eq.(\ref{eq:PWsol}) is not the solution of the original 
differential equation Eq.(\ref{eq:PW-dynamical2})}, irrespective of the parameters $T_0$ and $R_*$, i.e. 
independent of the initial conditions.  
  
To see this conclusion,  let us first define $B(T)$ to be $-1$ times 
the R.H.S. of Eq.(\ref{eq:PW-special2}), 
\begin{equation}
B(T):= (-{R_0}') \left[
                  \frac{\Delta}{R_0} + \frac{(-b') (1- {R_0}') }{2 (\Delta -1) }   ( \Delta -  {R_0}') 
                \right] .
\label{eq:B(T)}
\end{equation}
 What we   show below is that $B(T)$ varies (actually {\it diverges}) along the 
collapsing solution Eq.(\ref{eq:PWsol}) so that the starting assumption 
$\b$=constant (Eq.(\ref{eq:PW-special2})) is incompatible 
with the  dynamical equation 
Eq.(\ref{eq:PW-special1}). Causing such self-inconsistency, then,  Eq.(\ref{eq:PWsol}) is not the 
solution of the original dynamical equation, Eq.(\ref{eq:PW-dynamical2}). 

Now we prove  
\begin{description}
\item{\bf{Theorem 2}} \\
{\it
The quantity $B (T) $ diverges, and does not remain constant 
along the collapsing solution  Eq.(\ref{eq:PWsol}).  
}
\subitem{\it Proof:}\\ 
Let us look  at the definition of $B(T)$ (Eq.(\ref{eq:B(T)})). 
Both of the terms in the brackets $[\ \ \ ]$  are positive, 
since $ -1 < {R_0}' <0 $, $b' <0 $ and $\Delta > 1$. 
Thus $B(T) >0$ and 
accordingly, Eq.(\ref{eq:PW-special2}) implies 
that $\b <0$.  
We then get the estimation 
\[
B (T) > (-{R_0}') \frac{\Delta}{R_0} > \frac{(-{R_0}')}{R_0}\ \ ,
\]
so that 
\[
\lim_{T \rightarrow T^*} B(T) \geq 
\lim_{T \rightarrow T^*} \frac{(-{R_0}')}{R_0} \ \ ,  
\]
where $T^*$ is the shell-collapsing time, 
$\lim_{T \rightarrow T^*} R_0(T) =0$. \\
\quad \underline{ In the case $\lim_{T \rightarrow T^*} {R_0}'(T) \neq 0$}, 
the above estimation implies that $B(T) \rightarrow \infty $ as 
$T \rightarrow T^*$. \\
\quad \underline{ In the case   $\lim_{T \rightarrow T^*} {R_0}'(T) =0$}, 
the estimation further goes as
\begin{eqnarray*}
&& \lim_{T \rightarrow T^*} B(T) 
        \geq   \lim_{T \rightarrow T^*} \frac{(-{R_0}')}{R_0} \\
       && \ \ \ =\lim_{T \rightarrow T^*}  \frac{(-{R_0}'')}{{R_0}'}\ \ ,
\end{eqnarray*}
where  de L'Hopital's theorem has been applied in the last step.
With the help of Eq.(\ref{eq:PW-special1}), then, we get  
\begin{eqnarray*}
&& \lim_{T \rightarrow T^*} B(T) \geq   (-\b) \lim_{T \rightarrow T^*}   
                                            \frac{1-{{R_0}'(T)}^2}{{{R_0}'(T)}^2}  \\
       && \ \ \  > 
         \frac{1}{2} (-\b) \lim_{T \rightarrow T^*} \frac{1}{{{R_0}'(T)}^2}  = \infty 
         \ \ .
\end{eqnarray*}
Thus we get $\lim_{T \rightarrow T^*} B(T)= \infty$.   
\fbox{}
\end{description}

The situation is the same  for our correct equation, Eq.(\ref{eq:result1}) 
(with $P_0 >0$, $=0$ or $<0$), too. 
Just as Eqs.(\ref{eq:PW-special1}) and (\ref{eq:PW-special2}), 
we can ask whether the collapsing solution is allowed for the equation
\begin{eqnarray}
 \ddot{\rho}_+&=& \tilde{\b} \frac{1-{{\dot{\rho}}_+}^2}{\dot{\rho}_+}   \ \ , \label{eq:my-special1} \\
 \tilde{\b} :&=& \dot{\rho}_+ \big[
                        \frac{\Delta}{\rho_+} + \frac{\g' \dot{\rho}_+} {(\Delta -1) (1+ \dot{\rho}_+ ) }  
                         \{(\Delta +1) \dot{\rho}_+  + 1 \} \nonumber \\  
                       &&  +\frac{\Delta}{\Delta-1}(1-\dot{\rho_+}^2 )^{1/2} e^{\g_+} P_0 \big]\ \ ,  
\label{eq:my-special2}                
\end{eqnarray}
with $\tilde{\b}$ being a {\it constant}; $\Delta := \left[ (- \dot{\rho}_+)^2 + e^{2\g_+}\{1-(- \dot{\rho}_+)^2\}   \right] ^{1/2}$. 
Equation (\ref{eq:my-special1}) itself has a formal solution, which is 
just the same as Eqs.(\ref{eq:PWsol0}) and (\ref{eq:PWsol}). Though more care is required, 
essentially the same argument as above follows and it turns out that the formal solution 
just like Eq.(\ref{eq:PWsol}) is not the solution of the original dynamical equation Eq.(\ref{eq:result1}). 
To show this fact, we first introduce some symbols. 

\begin{description}
\item{\bf{Definition}} \\
\item{(a)}  
 Let $\tB $  be $-1$ times the R.H.S. of Eq.(\ref{eq:my-special2});
\begin{eqnarray}
&& \tB (t):= V \times  \nonumber \\
&&\ \times \big[\frac{\Delta}{\rho_+} 
                   + \frac{\g' V} 
                   {(\Delta -1) (1-V ) }  
                         [(\Delta +1)V  - 1] \nonumber \\
 && \  +\frac{\Delta}{\Delta -1}(1-V^2)^{1/2} e^{\g_+} P_0  \big] \ \ , 
\label{eq:tB(t)}
\end{eqnarray}
where $V:=- \dot{\rho}_+ $ ($ 0 < V < 1$).  
\item{(b)}
Let $\tD$ and $\tE$ be the second and the third terms in the brackets $[\ \ \ ]$ 
in Eq.(\ref{eq:tB(t)}), respectively;
\begin{eqnarray*} 
\tD(t) :&=& \frac{\g' V }{(\Delta -1) (1-V) } [(\Delta +1) V  - 1] \ \ . \\
\tE(t):&=& \frac{\Delta}{\Delta -1}(1-V^2)^{1/2} e^{\g_+} P_0 \ \ .
\end{eqnarray*}
\item{(c)} Let  $f(V):= (\Delta +1) V - 1$. 
The sign of $f(V)$ determines the sign of $\tD(t)$. 
As is already discussed in 
the previous section (after {\it Lemma 1}), the function $f(V)$ 
 is  monotonic, increasing, continuous on $[0, \ 1]$, with $f(0)=-1$ and $f(1)=1 $, so that $f(V)$ has only one zero, say $V^*$, 
in $(0,1)$. When  $\tilde{D}$ is looked as a function of $V$, then,  
$\tilde{D} \leq 0 $ for $V \in (0, V^*]$ and 
$\tilde{D} > 0 $ for $V \in (V^*, 1)$. 
\item{(d)} 
Let $t^*$ be the shell-collapsing time (which is assumed to exist 
in the present context), i.e.  
$t^*$ is the time such that $\lim_{t\rightarrow t^*} \rho_+(t) =0$. 
(It does not matter whether $t^*$ is finite or infinite in the 
argument below.) 
\end{description}

 We now prove a theorem ({\it Theorem 3} below) 
corresponding to {\it Theorem 2} step by step.

We first note 
\begin{description}
\item{\bf{Lemma 3}} \\
{\it $|\tE|/ \tD \rightarrow 0 $ as $V \rightarrow 1$.}
\subitem{\it Proof:}\\ 
When $V$ is close to 1, setting $V=1-\d$ ($\d >0$),  then $\Delta = 1 + (e^{\g_+} -1) \d +O(\d ^2) $, 
and one finds 
$1/|\tD| = O(\d^2) $ and $1/|\tE| = O(\d^{1/2})$, so that  
$|\tE|/\tD = O(\d^{3/2})$. Then the claim follows.  
\fbox{}
\end{description}

Now let us confirm 
\begin{description}
\item{\bf{Lemma 4}} \\
{\it
It follows that  $\lim_{t \rightarrow t^*} V (t) < 1 $.  
 }
\subitem{\it Proof:}\\ 
Studying the solution of Eq.(\ref{eq:my-special1}) (cf. Eq.(\ref{eq:PWsol0})), 
we see $V < 1 $ (i.e. $\dot{\rho}_+ (t) > -1 $). 
Suppose $\lim_{t \rightarrow t^*} V (t) = 1$. Since $f(V)$ is continuous 
on $[0, \ 1]$, then,   
it means $\lim_{t \rightarrow t^*} f (V(t))=1$, so that 
$\tD(t) \rightarrow \infty$ as well as 
 $\frac{\Delta}{\rho_+} \rightarrow \infty$ as 
$t \rightarrow t^*$ (note  $\Delta > 1$). 
Considering the limiting behavior of  $\tB(t)= V(t) [\frac{\Delta}{\rho_+} + \tD(t)(1 + \tE(t)/\tD(t) ) ] $ 
as $t \rightarrow t^*$, {\it Lemma 3} implies  
$\tB(t) \rightarrow \infty $ as $t \rightarrow t^*$. 
Since $\tilde{\b}$ ($= - \tB(t)$) is a constant by assumption, this should not happen.
 Thus $\lim_{t \rightarrow t^*} V (t) \neq 1$. 
\fbox{}
\end{description}
\begin{description}
\item{\bf Corollary of Lemma 4} \\
{\it
The quantities $\tD(t)$ and $\tE(t)$ are bounded, 
$|\tD (t)| < \infty$ and $|\tE (t)| < \infty$. Furthermore, 
$\lim_{t \rightarrow t^*} |\tD (t)| < \infty$ and 
$\lim_{t \rightarrow t^*} |\tE (t)| < \infty$. 
}
\subitem{\it Proof:}\\
We note that $\tD$ and $\tE$ become unbounded only when $V \rightarrow 1$ and 
that $\tD$ and $\tE$ are continuous as functions of $V$ .  
Now, the solution of Eq.(\ref{eq:my-special1}) (cf. Eq.(\ref{eq:PWsol0})) 
implies $V < 1 $. Furthermore {\it Lemma 4} says  $\lim_{t \rightarrow t^*} V (t) < 1 $.
Then the claim follows.  
\fbox{}
\end{description}

Then we can show  
\begin{description}
\item{\bf{Lemma 5}} \\
{\it
The constant parameter $\tilde{\b}$ should be set 
negative, $\tilde{\b} < 0. $}
\subitem{\it Proof:}\\ 
We note that 
$\lim_{t\rightarrow t^*} |\tD(t)| < \infty$ and 
$\lim_{t\rightarrow t^*} |\tE(t)| < \infty$ due to {\it Corollary of Lemma 4}. 
On the other hand,  $\lim_{t\rightarrow t^*} \frac{\Delta}{\rho_+} = \infty$. Then   
one can  find appropriate 
$t_1$ and $t_2$ ($t_1 < t_2 < t^*$) such that   
$\frac{\Delta}{\rho_+}  > |\tilde{D}(t)|+ |\tilde{E}(t)|   $ on $I_1:=(t_1, \ t_2)$. 
Thus $\tB(t) > 0$ on $I_1$. 
 Since $\tilde{\b}$ ($=-\tB(t)$)  is a constant by assumption, 
 this means that $\tilde{\b}$ should be set negative from 
 the outset.
\fbox{}
\end{description}

Finally we  prove 
\begin{description}
\item{\bf{Theorem 3}} \\
{\it
The quantity $\tB (T) $ diverges, and does not remain constant 
along the collapsing solution (corresponding to Eq.(\ref{eq:PWsol})).  
}
\subitem{\it Proof:}\\ 
Let us set $\tC(t):=\tB(t) - (- \dot{\rho_+})\big[\tD(t) + \tE(t)\big]$.
We note $\tC(t)= (-\dot{\rho}_+) \frac{\Delta}{\rho_+}$ $(> 0)$
 due to Eq.(\ref{eq:tB(t)}). Then it follows 
$\tC (t) >  \frac{(-\dot{\rho}_+(t))}{\rho_+ (t)} $.  
Noting $\tilde{\b} < 0$ ({\it Lemma 5}), then,  
 exactly the same argument  as the proof of  
{\it Theorem 2} can be applied,  and  
one concludes $\lim_{t \rightarrow t^*} \tC(t) = \infty$. 
On the other hand, it follows  
\begin{eqnarray*}
&&  \lim_{t \rightarrow t^*} \tC(t) \leq  \lim_{t \rightarrow t^*}|\tB(t)|  \\
&& \quad + \lim_{t \rightarrow t^*}|(- \dot{\rho}_+)\tD (t)| 
    + \lim_{t \rightarrow t^*}|(- \dot{\rho}_+)\tE (t)| \ \ . 
\end{eqnarray*}
Due to  {\it Corollary of Lemma 4}, then, 
it follows $\lim_{t \rightarrow t^*} |\tB(t)|=\infty $.  
\fbox{}
\end{description}

We conclude that there is no relevant solution indicating the singularity formation in the present model.

\section{
\label{sec:5}
Summary}

In this paper, 
we have studied the dynamics of a cylindrical shell 
in the spacetime cylindrical symmetry. 

We have started with constructing  a general framework for analyzing 
a cylindrical  spacetime and a shell in it, which might be useful for future investigations. 
Based on the framework, we have investigated 
a cylindrical shell-collapse model which accompanies  the out-going
radiation of gravitational waves and massless particles.  
This model  had been  introduced by Pereira and Wang~\cite{PeWa,PeWaE}, but its proper analysis 
had been awaited. 
This model could be interpreted as a thin shell filled with radiation. 
In connection with the cosmic-censorship hypothesis, we are mostly interested in the collapse of ``normal" matter, 
so that the weak  energy condition has been assumed outside the shell.
   
It has been proved that, as far as the weak energy condition is satisfied outside the shell,  
the shell bounces due to the rotational-pressure effect. After the bounce,  
it  continues  to expand without re-contraction when the pressure of the shell in the z-direction $p_z$ 
satisfies $p_z \geq0$, while in the case $p_z < 0$, the behavior after the bounce can be more complicated.
However in either case, the shell never reaches the zero-radius configuration and it escapes from the 
line-like singularity formation. 

Just as the case of a  shell filled with counter-rotating dust particles considered by Apocatopolas and Thorne~\cite{ApThorne}, 
the present case also shows  a bouncing due to the effect of rotational pressure, while  
the eternal expansion after the bounce (when $p_z \geq 0$) is a unique feature of the present model. 
We have also performed  numerical investigations which reveal 
explicit behaviors of the shell.

\begin{acknowledgments}
The author would like to thank S. Jhingan for valuable discussions 
at the beginning stage of  this work. He also thanks H. Sakuragi for 
valuable discussions on numerical investigations. Important part of  work has been completed during 
the author's stay at Institute of Cosmology, Tufts University. He 
is  grateful for its nice hospitality. In particular he would like to 
thank A. Vilenkin for helpful suggestions on the interpretation 
of the results. 
This work has been supported by the Japan  Ministry of Education, Culture, 
Sports, Science and Technology  with the grant \#14740162.
\end{acknowledgments}



\appendix

\section{
\label{app:A}
Summary of fundamental geometrical quantities for cylindrical spacetimes}

Based on the metric Eq.(\ref{eq:ds}), it is convenient to 
introduce the 1-forms as  Eq.(\ref{eq:cartan}).
Taking the exterior derivative of $\thh^A$'s, we get 
\begin{eqnarray}
d\thh^0 &=& -\frac{T'}{TR}\ \thh^0 \^ \thh^1\ \ , \ \ 
d\thh^1 = \frac{\dot{R}}{TR}\ \thh^0 \^ \thh^1\ \ , \nonumber  \\
d\thh^2 &=& \frac{\dot{Z}}{TZ}\ \thh^0 \^ \thh^2 + 
           \frac{Z'}{RZ}\ \thh^1 \^ \thh^2 \ \ ,  \nonumber \\ 
d\thh^3 &=& \frac{\dot{\Phi}}{T \Phi}\ \thh^0 \^ \thh^3 + 
           \frac{\Phi'}{R \Phi}\ \thh^1 \^ \thh^3 .  
\label{eq:dtheta}           
\end{eqnarray}
Here ``$\: \cdot{}\: $"  indicates the  partial derivative 
w.r.t.   $t$, while ``$\: {}'\: $" means the same w.r.t. $r$.  
With the help of the first Cartan structure-equation $d\thh^A + {\om^A}_B\ \^ \thh^B =0$,
we thus get 
\begin{eqnarray}
{\om^0}_1 &=& \frac{T'}{TR}\ \thh^0 + \frac{\dot{R}}{TR}\ \thh^1 \ \ , \ \ 
{\om^0}_2 = \frac{\dot{Z}}{TZ}\ \thh^2 \ \ ,  \nonumber  \\
{\om^0}_3 &=& \frac{\dot{\Phi}}{T \Phi}\ \thh^3 \ \ , \ \ 
{\om^1}_2 = - \frac{Z'}{RZ}\ \thh^2 \ \ , \ \ \nonumber \\
{\om^1}_3 &=& - \frac{\Phi'}{R \Phi}\ \thh^3 \ \ , \ \ 
{\om^2}_3 = 0 \ \ . 
\label{eq:omega}
\end{eqnarray}
On account of the relation, ${\om^A}_B = {\G^A}_{BC}\ \thh^C$, 
we also get~\footnote{
We recall that 
$\{ \eh_A \}_{A=0,1,2,3}$ forms the {\it non-coordinate bases} and 
$\G^A_{BC}$'s are not tensors , so that there is no surprise in that  
 ${\G^0}_{01}=0$ $(\neq {\G^0}_{10})$.} 
\begin{eqnarray}
{\G^0}_{10} &=& \frac{T'}{TR}  \ \ , \ \
{\G^0}_{11} = \frac{\dot{R}}{TR} = \frac{1}{XTR} (\Rdot - \rhodot R')  \ \ ,  \nonumber \\ 
{\G^0}_{22} &=& \frac{\dot{Z}}{TZ} = \frac{1}{XTZ} (\Zdot - \rhodot Z')  \ \ ,  \nonumber \\ 
{\G^0}_{33} &=& \frac{\dot{\Phi}}{T\Phi} 
                      = \frac{1}{XT\Phi} (\Phidot - \rhodot \Phi')  \ \ , \nonumber \\  
{\G^1}_{22} &=& - \frac{Z'}{RZ}   \ \ ,  \ \  
{\G^1}_{33} = - \frac{\Phi'}{R\Phi}   \ \ ,  \nonumber \\
      {\rm others} &=& 0\ .  
\label{eq:Gamma}      
\end{eqnarray}
Here `` ${}^\circ $ " indicates the derivative w.r.t.  
the proper-time $\tau$ of an observer on the shell, and  $X:=dt/d\t$; 
they are  evaluated on the shell in consideration.

Taking the exterior-derivatives of ${\om^A}_B$'s,  
we get 
\begin{eqnarray}
{d\om^0}_1 &=& \frac{1}{TR}\ 
                 \left\{ \Bigg( \frac{\dot{R}}{T}\dot{\Bigg)\ }
                    - \left( \frac{T'}{R}\right)' \right\}
                \thh^0 \^ \thh^1  \ \ , \nonumber \\
{d\om^0}_2 &=& \frac{1}{TZ}\ \Bigg( \frac{\dot{Z}}{T}\dot{\Bigg)\ } 
                     \thh^0 \^ \thh^2 
                  +   \frac{1}{RZ}\ \left( \frac{\dot{Z}}{T}\right)' 
                     \thh^1 \^ \thh^2 
                     \ \ , \nonumber \\
{d\om^0}_3 &=& \frac{1}{T\Phi}\ \Bigg( \frac{\dot{\Phi}}{T}\dot{\Bigg)\ } 
                     \thh^0 \^ \thh^3 
                  +   \frac{1}{R\Phi}\ \left( \frac{\dot{\Phi}}{T}\right)' 
                     \thh^1 \^ \thh^3 
                     \ \ , \nonumber \\
{d\om^1}_2 &=& - \frac{1}{TZ}\ \Bigg( \frac{Z'}{R}\dot{\Bigg)\ } 
                     \thh^0 \^ \thh^2 
                  -   \frac{1}{RZ}\ \left( \frac{Z'}{R}\right)' 
                     \thh^1 \^ \thh^2 
                     \ \ , \nonumber \\
{d\om^1}_3 &=& - \frac{1}{T\Phi}\ \Bigg( \frac{\Phi'}{R}\dot{\Bigg)\ } 
                     \thh^0 \^ \thh^3 
                  -   \frac{1}{R \Phi}\ \left( \frac{\Phi'}{R}\right)' 
                     \thh^1 \^ \thh^3 
                     \ \ , \nonumber \\
{d\om^2}_3 &=& 0 \ \ . 
\label{eq:domega}
\end{eqnarray}
Thus, by the second Cartan structure-equation ${\Omega^A}_B = d{\om^A}_B + {\om^A}_C  \^ {\om^C}_B $, 
we find
\begin{widetext}
\begin{eqnarray}
{\Om^0}_1 &=& \frac{1}{TR}\ 
                 \left\{ \Bigg( \frac{\dot{R}}{T} \dot{\Bigg)\ }  
                    - \left( \frac{T'}{R}\right)' \right\}
                \thh^0 \^ \thh^1  \ \ , \ \ \nonumber \\
{\Om^0}_2 &=& \left\{
                        \frac{1}{TZ}\ \Bigg( \frac{\dot{Z}}{T}\dot{\Bigg)\ } 
                        - \frac{1}{R^2}\frac{T'Z'}{TZ}  
                 \right\}       \thh^0 \^ \thh^2  
                +  \left\{
                     \frac{1}{RZ}\ \left( \frac{\dot{Z}}{T}\right)'
                     -\frac{1}{TR}\frac{\dot{R}Z'}{RZ} 
                 \right\}    \thh^1 \^ \thh^2 
                     \ \ , \nonumber \\
{\Om^0}_3 &=& \left\{
                    \frac{1}{T\Phi}\ \Bigg( \frac{\dot{\Phi}}{T}\dot{\Bigg)\ } 
                        - \frac{1}{R^2}\frac{T'\Phi'}{T\Phi}  
                 \right\}       \thh^0 \^ \thh^3 
               +  \left\{
                     \frac{1}{R\Phi}\ \left( \frac{\dot{\Phi}}{T}\right)'
                     -\frac{1}{TR}\frac{\dot{R}\Phi'}{R\Phi} 
                 \right\}    \thh^1 \^ \thh^3
                     \ \ , \nonumber \\                     
{\Om^1}_2 &=& - \left\{
                    \frac{1}{TZ}\ \Bigg( \frac{Z'}{R}\dot{\Bigg)\ } 
                        - \frac{1}{TR}\frac{T'\dot{Z}}{TZ}  
                 \right\}       \thh^0 \^ \thh^2 
                  -  \left\{
                     \frac{1}{RZ}\ \left( \frac{Z'}{R}\right)'
                     -\frac{1}{T^2}\frac{\dot{R}\dot{Z}}{RZ} 
                 \right\}    \thh^1 \^ \thh^2
                     \ \ , \nonumber \\                     
{\Om^1}_3 &=& - \left\{
                    \frac{1}{T\Phi}\ \Bigg( \frac{\Phi'}{R}\dot{\Bigg)\ } 
                        - \frac{1}{TR}\frac{T'\dot{\Phi}}{T\Phi}  
                 \right\}       \thh^0 \^ \thh^3 
               -  \left\{
                     \frac{1}{R\Phi}\ \left( \frac{\Phi'}{R}\right)'
                     -\frac{1}{T^2}\frac{\dot{R}\dot{\Phi}}{R\Phi} 
                 \right\}    \thh^1 \^ \thh^3
                     \ \ , \nonumber \\                     
{\Om^2}_3 &=& \left( 
                 \frac{1}{T^2}\frac{\dot{Z}\dot{\Phi}}{Z\Phi}
               -  \frac{1}{R^2}\frac{Z' \Phi'}{Z\Phi}
              \right) \thh^2 \^ \thh^3 \ \ . 
\label{eq:Omega}              
\end{eqnarray}
Using  the relation ${\Omega^A}_B = \frac{1}{2} {\bf R}^A_{\: BCD}\: \thh^C\: \^ \thh^D$, 
we obtain the Riemann curvature w.r.t. the frame $\{ \eh_A \}$,  
\begin{eqnarray}
{{\bf R}^0}_{101}&=& \frac{1}{TR}\ 
                 \left\{ \Bigg( \frac{\dot{R}}{T}\dot{\Bigg)\ }
                    - \left( \frac{T'}{R}\right)' \right\}   \ \ , \ \ 
{{\bf R}^0}_{202} = \frac{1}{TZ}\ \Bigg( \frac{\dot{Z}}{T}\dot{\Bigg)\ } 
                        - \frac{1}{R^2}\frac{T'Z'}{TZ}   \ \ , \ \
{{\bf R}^0}_{212} =  \frac{1}{RZ}\ \left( \frac{\dot{Z}}{T}\right)'
                     -\frac{1}{TR}\frac{\dot{R}Z'}{RZ}   \ \ , \ \ \nonumber \\
{{\bf R}^0}_{303} &=& \frac{1}{T\Phi}\ \Bigg( \frac{\dot{\Phi}}{T}\dot{\Bigg)\ } 
                    - \frac{1}{R^2}\frac{T'\Phi'}{T\Phi}     \ \ , \ \ 
{{\bf R}^0}_{313} =    \frac{1}{R\Phi}\ \left( \frac{\dot{\Phi}}{T}\right)'
                     -\frac{1}{TR}\frac{\dot{R}\Phi'}{R\Phi}  \ \ , \ \ 
{{\bf R}^1}_{202} = -\frac{1}{TZ}\ \Bigg( \frac{Z'}{R}\dot{\Bigg)\ } 
                        + \frac{1}{TR}\frac{T'\dot{Z}}{TZ}  
                     \ \ , \nonumber \\                     
{{\bf R}^1}_{212} &=&  -\frac{1}{RZ}\ \left( \frac{Z'}{R}\right)'
                     +\frac{1}{T^2}\frac{\dot{R}\dot{Z}}{RZ} \ \ , \ \                  
{{\bf R}^1}_{303} = - \frac{1}{T\Phi}\ \Bigg( \frac{\Phi'}{R}\dot{\Bigg)\ } 
                        + \frac{1}{TR}\frac{T'\dot{\Phi}}{T\Phi}    
                     \ \ , \ \                     
{{\bf R}^1}_{313} =  - \frac{1}{R\Phi}\ \left( \frac{\Phi'}{R}\right)'
                     + \frac{1}{T^2}\frac{\dot{R}\dot{\Phi}}{R\Phi} 
                     \ \ , \ \ \nonumber        \\               
{{\bf R}^2}_{323} &=& \frac{1}{T^2}\frac{\dot{Z}\dot{\Phi}}{Z\Phi}
               -  \frac{1}{R^2}\frac{Z' \Phi'}{Z\Phi} \ \ , \ \ 
      {\rm  others} = 0 \ \ . 
\label{eq:Riemann}               
\end{eqnarray}
Thus, the Ricci curvature in the frame $\{\eh_A \}$ becomes
\begin{eqnarray}
{\bf R}_{00} &=& -\frac{1}{T}
                 \left\{
                 \frac{1}{R}\Bigg(\frac{\dot{R}}{T}\dot{\Bigg)\ }
                + \frac{1}{Z}\Bigg(\frac{\dot{Z}}{T}\dot{\Bigg)\ }
                + \frac{1}{\Phi}\Bigg(\frac{\dot{\Phi}}{T}\dot{\Bigg)\ }
                 \right\}  
            +\frac{1}{TR}\left(\frac{T'}{R} \right)'
           +\frac{1}{R^2}\frac{T'}{T}
                \left( \frac{Z'}{Z} + \frac{\Phi'}{\Phi} 
                \right)  \ \ , \nonumber \\ 
{\bf R}_{01} &=& {\bf R}_{10}= -\frac{1}{R}
                 \left\{
                 \frac{1}{Z}\left(\frac{\dot{Z}}{T}\right)'
                + \frac{1}{\Phi}\left(\frac{\dot{\Phi}}{T}\right)'
                 \right\}  
             +\frac{1}{TR}\frac{\dot{R}}{R}
                \left( \frac{Z'}{Z} + \frac{\Phi'}{\Phi} 
                \right)  \ \ , \nonumber \\ 
{\bf R}_{11} &=&  \frac{1}{TR}\Bigg(\frac{\dot{R}}{T}\dot{\Bigg)\ }
           +\frac{1}{T^2}\frac{\dot{R}}{R}
                   \left(\frac{\dot{Z}}{Z} + \frac{\dot{\Phi}}{\Phi}  \right) 
           -\frac{1}{R}\left\{ \frac{1}{T}\left( \frac{T'}{R} \right)'
                              +\frac{1}{Z}\left( \frac{Z'}{R} \right)'
                              +\frac{1}{\Phi}\left( \frac{\Phi'}{R} \right)'
                       \right\}     \ \ , \nonumber \\
{\bf R}_{22} &=&  \frac{1}{TZ}\Bigg(\frac{\dot{Z}}{T}\dot{\Bigg)\ }
           +\frac{1}{T^2}\frac{\dot{Z}}{Z}
                   \left(\frac{\dot{R}}{R} + \frac{\dot{\Phi}}{\Phi}  \right) 
           -\frac{1}{RZ}\left( \frac{Z'}{R} \right)'
           -\frac{1}{R^2}\frac{Z'}{Z}
               \left( \frac{T'}{T} + \frac{\Phi'}{\Phi} \right) 
                                                    \ \ , \nonumber \\
{\bf R}_{33} &=&  \frac{1}{T\Phi}\Bigg(\frac{\dot{\Phi}}{T}\dot{\Bigg)\ }
           +\frac{1}{T^2}\frac{\dot{\Phi}}{\Phi}
                   \left(\frac{\dot{R}}{R} + \frac{\dot{Z}}{Z}  \right)  
           -\frac{1}{R\Phi}\left( \frac{\Phi'}{R} \right)'
           -\frac{1}{R^2}\frac{\Phi'}{\Phi}
               \left( \frac{T'}{T}+ \frac{Z'}{Z} \right) \ \ ,  \\
           {\rm others}&=& 0 \ \ . \nonumber 
\label{eq:Ricci}               
\end{eqnarray}
Finally,  the scalar curvature becomes,  independently of the frame, 
\begin{eqnarray}
 {\bf R} &=& \frac{2}{T}\left\{
                 \frac{1}{R}\Bigg(\frac{\dot{R}}{T}\dot{\Bigg)\ }
                + \frac{1}{Z}\Bigg(\frac{\dot{Z}}{T}\dot{\Bigg)\ }
                + \frac{1}{\Phi}\Bigg(\frac{\dot{\Phi}}{T}\dot{\Bigg)\ }
                 \right\}
                +\frac{2}{T^2}
                   \left(
                   \frac{\dot{R}\dot{Z}}{RZ} + 
                   \frac{\dot{R}\dot{\Phi}}{R\Phi} + 
                   \frac{\dot{Z} \dot{\Phi}}{Z\Phi} 
                   \right) \nonumber  \\
   & &   -\frac{2}{R}
                       \left\{ \frac{1}{T}\left( \frac{T'}{R} \right)'
                             +\frac{1}{Z}\left( \frac{Z'}{R} \right)'
                             +\frac{1}{\Phi}\left( \frac{\Phi'}{R} \right)'
                       \right\}
                   -\frac{2}{R^2} 
                      \left(
                         \frac{T'Z'}{TZ} + \frac{T'\Phi'}{T\Phi} 
                             +\frac{Z'\Phi'}{Z \Phi}
                      \right) . \ \ \ 
\label{eq:R}                      
\end{eqnarray}
\end{widetext}

\section{
\label{app:B}
Non-coordinate bases}

Throughout this paper, we always consider a cylindrical spacetime defined by the metric 
Eq.(\ref{eq:ds}). The coordinates $\{ x^\m \}:=\{ t,\ r,\ z,\ \phi \}$ define the coordinate bases and 
 their dual 1-forms,
\begin{eqnarray}
e_\m :=\del_\m\ \ , \ \ \th^\m := dx^\m \ \ .
\end{eqnarray}
(Here the Greek letters $\m, \n, \cdots$  indicate  
the indices w.r.t.  the coordinates $\{ t,\ r,\ z,\ \phi \}$.)

On the other hand, a cylindrical shell embedded in the spacetime 
naturally  defines a  local ortho-normal frame $\{ \eh_\a \}$ (see Sec.\ref{sec:2-1}), 
\begin{eqnarray}
{\eh_n} &=& \nh^\m =\left(\frac{R}{T} \rhodot,\ \frac{T}{R} X,\ 0,\ 0 
                  \right)_{(trz\phi)}\ \ ,   \nonumber \\
{\eh_\t} &=& \partial_\t =  
           (X,\ \rhodot,\ 0,\ 0 )_{(trz\phi)}\ \ ,    \\
{\eh_z} &=&  
           \left(0,\ 0,\ 1/Z,\ 0 
                  \right)_{(trz\phi)}\ \ ,   \nonumber  \\
{\eh_\phi} &=&  \left(0,\ 0,\ 0,\ 1/\Phi 
                  \right)_{(trz\phi)}\ \ ,  
\label{eq:ehat}                                           
\end{eqnarray}
where the  symbol $\eh$ implies the  normalized frame-vector.  
(The first few Greek letters   
$\a$, $\b$, $\g$, $\cdots$ are  used for the indices w.r.t.  
the above ortho-normal frame, and they  take the values $n$, $\t$, 
$z$ or $\phi$.) 
A set of 1-forms $\{\thh^\a \}$ dual to $\{ \eh_\a \}$ is also introduced.

As the third set of frames, 
the metric given in Eq.(\ref{eq:ds}) defines a natural set of 
1-forms $\{ \thh^A  \}$, 
\begin{eqnarray}
\thh^0 &=& T(t,r) dt \ \ , \ \ 
\thh^1 = R(t,r) dr \ \ , \nonumber \\
\thh^2 &=& Z(t,r) dz \ \ , \ \ 
\thh^3 = \Phi(t,r) d\phi \ \ ,
\label{eq:frame3}
\end{eqnarray}
along with their dual ortho-normal bases $\{ \eh_A \}$. 
(The capital Latin letters $A$, $B$, $C$, $\cdots$ 
 indicate the indices w.r.t.  
the frame  determined by  Eq.(\ref{eq:frame3}),  and they  take the values $0-3$.)

Now we summarize the relations among $\{ e_\m  \}$, $\{ \eh_\a  \}$ and  $\{ \eh_A  \}$,  and 
those among $\{ \th^\m \}$, $\{ \thh^\a \}$ and  $\{ \thh^A \}$. 

Firstly,  Eq.(\ref{eq:ehat}) gives  the relation between $\{ e_\m  \}$ and  $\{ \eh_\a  \}$,  and those between 
$\{ \th^\m \}$ and $\{ \thh^\a \}$  
\begin{equation}
{\eh_\a} = {e_\m}\  {E^\m}_\a\ \ , \ \ 
{\thh^\a} = {E^\a}_\m\  {\th^\m} \ \ . 
\label{eq:e1}
\end{equation}
Here 
${E^\m}_\a$ and  ${E^\b}_\n$  defined below  are mutually inverse as matrices: 
\begin{eqnarray}
{E^\m}_\a 
&=& \left(
  \begin{array}{cccc}
     T^{-1} R \rho^\circ    & X         & 0     & 0         \\
     X T R^{-1}           & \rho^\circ   & 0     & 0          \\
      0                 & 0                  & Z^{-1}   & 0    \\
      0                 & 0                   & 0     & \Phi^{-1}     
  \end{array}
\right)
\ \ , \nonumber \\ 
{E^\a}_\m 
&=& \left(
  \begin{array}{cccc}
     -T R \rho^\circ  & XTR           & 0     & 0    \\
     XT^2          & -R^2 \rho^\circ   & 0     & 0    \\
      0          & 0                  & Z   & 0    \\
      0          & 0                   & 0     & \Phi     
  \end{array}
\right)\ \ .
\label{eq:E1}
\end{eqnarray}

Secondly, Eq.(\ref{eq:frame3}) gives 
  the relation between $\{ e_\m  \}$ and  $\{ \eh_A  \}$,  and those between 
$\{ \th^\m \}$ and $\{ \thh^A \}$ 
\begin{eqnarray}
{\thh^A} &=& {E^A}_\m\  {\th^\m}\ \ , \ \ 
          {\th^\m}= {E^\m}_A {\thh^A} \ \ , \nonumber \\
{\eh_A} &=& {e_\m} \  {E^\m}_A \ \ , \ \ 
          {e_\m} = {\eh_A} \ {E^A}_\m\ \ .
\label{eq:e2}
\end{eqnarray}
Here 
${E^A}_\m$ and  ${E^\m}_A$  defined below  are mutually inverse as matrices: 
\begin{eqnarray}
{E^A}_\m &=& diag (T, \ R,\ Z, \ \Phi )\ \ , \nonumber \\ 
{E^\m}_A &=& diag (1/T, \ 1/R,\ 1/Z, \ 1/\Phi )\ \ .
\label{eq:E2}
\end{eqnarray}
Finally, Eqs.(\ref{eq:e1}) and (\ref{eq:e2}) give 
  the relation between $\{ \eh_A  \}$ and  $\{ \eh_\a  \}$,  and those between 
$\{ \thh^A \}$ and $\{ \thh^\a \}$
\begin{equation}
{\eh_\a}={\eh_A} {E^A}_\a\ \ , \ \
{\thh^\a}={E^\a}_{A} {\thh^A}\ \ . 
\label{eq:e3}
\end{equation}
Here 
${E^A}_\a := {E^A}_\m {E^\m}_\a$ and  ${E^\a}_A := {E^\a}_\m {E^\m}_A$  
given below  are mutually inverse as matrices: 
\begin{eqnarray}
{E^A}_\a 
  &=& \left(
     \begin{array}{cccc}
      R \rho^\circ  & XT           & 0     & 0    \\
     XT          & R \rho^\circ    & 0     & 0    \\
      0          & 0            & 1     & 0    \\
      0          & 0            & 0     & 1     
  \end{array}
\right)
\ \ , \nonumber \\ 
{E^\a}_A  
    &=& \left(
     \begin{array}{cccc}
      -R \rho^\circ  & XT           & 0     & 0    \\
       XT          & -R \rho^\circ    & 0     & 0    \\
      0          & 0            & 1     & 0    \\
      0          & 0            & 0     & 1     
  \end{array}
\right) \ \ .
\label{eq:E3}
\end{eqnarray}

Now noting the relation~\cite{Nakahara},
\[
\nabla_{\eh_\a}{\eh_\b} = \G^\g_{\a\b} {\eh_\g} \ \ ,
\]
along with Eq.(\ref{eq:e3}), we get 
\begin{eqnarray}
\G^\g_{\a\b} &=& {E^A}_\a {E^B}_\b {E^\g}_C \G^C_{AB}
 + {E^A}_\a {E^\g}_B \nabla_{\eh_A}{E^B}_\b  \nonumber \\
           &=& {E^A}_\a {E^B}_\b {E^\g}_C \G^C_{AB}
 + {E^\m}_\a {E^\g}_B \partial_\m {E^B}_\b .
\label{eq:Gamma-rel}
\end{eqnarray}
Equation (\ref{eq:Gamma-rel}) is the basis for calculating the 
extrinsic curvature.

\section{
\label{app:C}
Derivation of the formula for $K_{e_\t e_\t}$}

Here derivations of the formula for the  extrinsic curvatures (Eqs.(\ref{eq:Kttfinal}) and (\ref{eq:Kttfinal2})) 
are purposefully  shown  in detail. 
This is to ensure   anyone interested in this topic 
 reproducing all the results presented in this paper without any difficulty  and to motivate him to 
 go ahead the point where this paper ends.

\begin{widetext}
In order to derive Eq.(\ref{eq:Kttfinal}), we  note that 
the first term on the R.H.S.  in Eq.(\ref{eq:Ktt}) becomes, 
with the help of Eqs.(\ref{eq:Gamma}) and (\ref{eq:E3}),   
\begin{eqnarray}
 -{E^A}_\t {E^B}_\t {E^n}_C \G^C_{AB} 
 &=& -{E^1}_\t {E^0}_\t {E^n}_0 \G^0_{10} -({E^1}_\t)^2 {E^n}_0 \G^0_{11} \nonumber \\
&& = (R\rhodot)^2 
        \left\{
           \frac{XT}{2R}(\ln T^2)' 
              +  \frac{R}{2XT} \rhodot   \left[(\ln R^2)^\circ - \rhodot (\ln R^2)'   \right]  
        \right\} \ \ .
\label{eq:calc1}
\end{eqnarray}

Similarly, with the help of Eqs.(\ref{eq:E1}) and  (\ref{eq:E3}), the second term on the R.H.S. in Eq.(\ref{eq:Ktt}) 
is calculated as 
\begin{eqnarray*}
 -{E^\m}_\t {E^n}_B \partial_\m {E^B}_\t 
&=&  - {E^n}_0 \left( {E^t}_\t \partial_t {E^0}_\t +  {E^r}_\t \partial_r {E^0}_\t  \right) 
   - {E^n}_1 \left( {E^t}_\t \partial_t {E^1}_\t +  {E^r}_\t \partial_r {E^1}_\t  \right) \nonumber  \\
&&=  R\rhodot (XT)^\circ - XT (R\rhodot)^\circ  \ \ .
\label{eq:calc2}
\end{eqnarray*}
\end{widetext}
The above expression can be further modified with the help of  the following formulas that can be 
derived straightforwardly;
\begin{eqnarray}
\Xdot &=& \frac{R^2}{XT^2} \rhodot\ \rhoddot - \frac{X}{2} (\ln T^2)^\circ 
                  + \frac{R^2}{2XT^2} {\rhodot}^{\: 2} (\ln R^2)^\circ \ \ , \nonumber \\
X' &=& -\frac{T^2X^3}{2}  (\ln T^2)'  
                  + \frac{X R^2}{2} {\rhodot}^{\: 2} (\ln R^2)'\ \ .                                                             
\end{eqnarray}

After some calculations we then get
\begin{eqnarray*}
K_{e_\t e_\t} &=& -\frac{R}{XT} \rhoddot + (\frac{XT}{2} - \frac{1}{XT}) R \rhodot (\ln R^2)^\circ \\ 
        &&+ \frac{XTR}{2}{\rhodot}^{\: 2} (\ln T^2)' - \frac{R^3}{2XT} {\rhodot}^{\: 4}(\ln R^2)' \\
         &=:& (I) + (II) +(III) + (IV) \ \ . \\
\end{eqnarray*}
The last two terms together yield
\begin{eqnarray*}
&& (III)+(IV) \\
&& = \frac{XTR}{2}{\rhodot}^{\: 2} 
                    \left[ XTR\  \del_n (\ln T^2/R^2)  -R^2 \rhodot\  (\ln T^2/R^2)^\circ    \right] \\
           &&\ \ \ \ \  + \frac{R}{2XT} {\rhodot}^{\: 2} 
                         \left[ XTR\  \del_n (\ln R^2) - R^2 \rhodot\  (\ln R^2)^\circ   
                          \right]         \\
           &&=:   (V_1)+(V_2)+(V_3)+(V_4)\ \ ,               
\end{eqnarray*}
where the formulas in  Eq.(\ref{eq:F-formula2}) have been used. 
Noting that 
$(II)+(V_4) = -\frac{R}{2XT}\rhodot\ (\ln R^2)^\circ$,  
thus,  the calculations go as
\begin{eqnarray*}
K_{e_\t e_\t}&=& (I) + (II) +(III) + (IV) \\
        &=& (I)+ \{(II)+(V_4)\} + (V_3) + \{ (V_1) + (V_2) \} \ \ ,
\end{eqnarray*}
yielding Eq.(\ref{eq:Kttfinal}). 

Now we derive Eq.(\ref{eq:Kttfinal2}).  
To start with,  let us  note the following formula derived straightforwardly from 
Eq.(\ref{eq:rhodot}):
\[
\dot{X} =  - \frac{X^3 T^2}{2} (\ln T^2 \dot{)\ } 
               + \frac{X^3 R^2}{2}\dot{\rho}^2 (\ln R^2\dot{)\ }
               + X^3 R^2\  \dot{\rho}\  \ddot{\rho} \ \ .  
\]
Taking the $t$-derivative of  the relation $\rhodot = X\dot{\rho}  $ (Eq.(\ref{eq:rhodot})), 
then, 
we  obtain, 
\begin{equation}
\ddot{\rho} = \frac{1}{X^4 T^2} \rhoddot + \frac{1}{2X^2} \rhodot (\ln T^2)^\circ
                  - \frac{R^2}{2X^4 T^2  } {\rhodot}^{\: 3} (\ln R^2)^\circ\ \ .
\label{eq:rhodotdot}
\end{equation}
We should  return to the basic equation Eq.(\ref{eq:Ktt}) along with 
 the calculations Eqs.(\ref{eq:calc1}) and (\ref{eq:calc2}). Noting the relations 
\begin{eqnarray*}
(XT)^\circ &=& X \{    (TX\dot{)\ } + \dot{\rho} (TX)'   \}   \\
                &=& \frac{XT}{2} (X^2T^2 -1) ( \ln ({R^2}/{T^2}) )^\circ  
                         + X^4 T R^2\  \dot{\rho}\ \ddot{\rho} \  \ ,    \\
(R\rhodot)^\circ &=& \Rdot \rhodot + R\rhoddot \\ 
                          &=& X^4 T^2 R \ddot{\rho}
                              + \frac{1}{2} X^3 T^2 R \ \dot{\rho}\ ( \ln ({R^2}/{T^2})  )^\circ \ \ , 
\end{eqnarray*}
we finally reach Eq.(\ref{eq:Kttfinal2}).

\section{
\label{app:D}
Detailed comparison with the formulas in the preceding paper}
\subsection{Extrinsic curvature}
Here we point out the discrepancy of 
the formulas reported in Ref.~\cite{PeWa} and  its correction Ref.\cite{PeWaE} 
with our formulas.

Let us note once again that their functions $f$, $g$, $h$, $l$, $R_0(T)$, ${R_0}'(T)$,  ${R_0}''(T)$ and $b(T-R)$ correspond to ours as 
\begin{eqnarray}
&& f:=T^2\  ,  \  g:=R^2\  ,   \ 
h:= Z^2 \  ,  \ l:= \Phi^2 \  , \nonumber  \\
&& R_0(T):=\rho(t)\  , \  {R_0}'(T):=\dot{\rho}(t)\  ,  \nonumber \\
&& {R_0}''(T):=\ddot{\rho}(t)  \ , \ b(T-R)=-2\g(t-r) \ \ .
\label{eq:notationmatchApp}
\end{eqnarray}

They show the expression  for the $(\t\t)$-component of the extrinsic curvature
 (see the first formula in Eq.(10) of Ref.\cite{PeWa}),
which we now denote $K^{(PW)}_{\t\t}$, as 
\begin{widetext}
\begin{eqnarray}
 K^{(PW)}_{\t\t} &=& -\frac{1}{2}\frac{(fg)^{1/2}}{  (f-g\  {{R_0}'(T)}^2)^{3/2}  } \times  \nonumber \\
     &&  \times  \left\{
-\frac{{f,}_R}{g }  + \left(  \frac{{f,}_T}{f } - 2\frac{{g,}_T}{g}  \right) {R_0}' (T) 
 + \left( 2 \frac{{f,}_R}{f } - \frac{{g,}_R}{g}  \right) ({R_0}' (T))^2
+ \frac{{g,}_T}{f} {{R_0}'(T)}^3  -2 {R_0}''(T)
    \right\}    \ \ .
\label{eq:K(PW)}
\end{eqnarray}
\end{widetext}
It is easily checked that their result  is  translated into  our notation as 
\begin{eqnarray}
  K^{(PW)}_{\t\t}
  &=& X^3 TR\  \ddot{\rho}  + \frac{XR\ \dot{\rho}}{2T} (\ln R^2 \dot{) \  } 
     + \frac{XT}{2R} (\ln T^2)' \nonumber \\
  && \ \ \   - \frac{1}{2}X^2 TR\  \dot{\rho} \ (\ln (T^2/R^2) )^\circ \ \ , 
\label{eq:K(PW2)}
\end{eqnarray}
where the $\t$-derivative in the last term is for making the expression  concise, and can be 
 expressed in terms of $``\cdot{}"$ and ``\ $'$ \ " if needed.
Comparing Eq.(\ref{eq:K(PW2)}) with Eq.(\ref{eq:Kttfinal2}), we 
realize, 
\begin{eqnarray}
&& K^{(PW)}_{\t\t} = -K_{e_\t e_\t} + \frac{X^3 T^3}{2R} 
                         \{  (\ln T^2)' + \frac{R^2}{T^2}\ \dot{\rho}\  (\ln R^2 \dot{)\  } \} \nonumber \\
              && \ \ \ =-K_{e_\t e_\t} + \frac{X^2T^2}{2} \del_n  \ln R^2 
                         + \frac{X^4 T^4}{2} \del_n \ln (T^2/R^2)  \nonumber \\
                            && \qquad \qquad \qquad  - \frac{X^3 T^3}{2} R \rhodot (\ln(T^2/R^2))^\circ \ \ .
\label{eq:K(PW)comparison}
\end{eqnarray}

Here we have made use of Eq.(\ref{eq:F-formula2}) to reach the last line in  Eq.(\ref{eq:K(PW)comparison}); 
\begin{eqnarray*}
&& (\ln T^2)' + \frac{R^2}{T^2}\ \dot{\rho}\ (\ln R^2 \dot{)\ \ } \\
&=& -R^2 \rhodot (\ln (T^2/R^2))^\circ  + XTR\  \del_n \ln (T^2/R^2) \\
 &&\ \ \  + \frac{R}{XT}\ \del_n \ln R^2 \ \ .
\end{eqnarray*}

On the other hand, their expressions for the $(zz)$-component and the $(\phi\phi)$-component 
of the extrinsic curvature, $K^{(PW)}_{zz}$ and $K^{(PW)}_{\phi \phi}$, 
(see the second and the third formulas in Eq.(10) of Ref.\cite{PeWa}) 
are related to 
our $K_{e_z e_z}$ and $K_{e_\phi e_\phi}$ as
\begin{eqnarray}
K^{(PW)}_{zz} &=&
-\frac{1}{2} \left(\frac{fg}{  f-g\  {{R_0}'(T)}^2 } \right)^{1/2}
   \left(  \frac{{h,}_R}{g } - \frac{{h,}_T}{f} {R_0}' (T)  \right)  \nonumber  \\
   &=& -Z^2 \del_n \ln Z  \nonumber \\
   &=& - ({E^{e_z}}_z)^2 K_{e_z e_z} \ \ . \\
   \label{eq:K(PW)zz}
K^{(PW)}_{\phi \phi} &=&
-\frac{1}{2} \left(\frac{fg}{  f-g\  {{R_0}'(T)}^2 } \right)^{1/2}
   \left(  \frac{{l,}_R}{g } - \frac{{l,}_T}{f} {R_0}' (T)  \right) \nonumber    \\
   &=& -\Phi^2 \del_n \ln \Phi  \nonumber   \\
   &=& - ({E^{e_\phi}}_\phi)^2 K_{e_\phi e_\phi} \ \ .
\label{eq:K(PW)phiphi}
\end{eqnarray}
It turns out  that $K^{(PW)}_{zz}$ and $K^{(PW)}_{\phi \phi}$ are totally consistent with our $K_{e_z e_z}$ and $K_{e_\phi e_\phi}$, taking into account that 
the former is based on  the coordinate frame $\{ \del_\m  \}_{\m=t,r,z,\phi}$, 
  while the latter is based on the ortho-normal frame $\{ \eh_\a \}_{\a=n,\t,z,\phi}$.
 [On the other hand,  $K^{(PW)}_{\t\t}$ is the $(\t \t)$-component (and not $(tt)$-component), and 
$\del_\t=\eh_\t$ is a member of  the ortho-normal frame $\{ \eh_\a \}_{\a=n,\t,z,\phi}$. 
Therefore  $K^{(PW)}_{\t\t}$ is what should correspond to our $K_{e_\t e_\t}$ (up to sign) 
without any further conversion factor.]

Looking at Eqs.(\ref{eq:K(PW)comparison})-(\ref{eq:K(PW)phiphi}), we realize
\begin{description}
\item{(0)} Their definition for the extrinsic curvature  and ours differ by the overall sign, which is just 
the matter of definition and is not essential. 
\item{(1)} The last three terms in the last line on the R.H.S. in Eq.(\ref{eq:K(PW)comparison}) are 
          the essential discrepancies 
          in the $(\t\t)$-component of the extrinsic curvature.
\item{(2)} There is no essential discrepancy regarding the $(zz)$-component and the $(\phi\phi)$-component 
of the extrinsic curvature.           
\end{description}

Their expression Eq.(\ref{eq:K(PW)}) can be modified so as to 
coincide with our result $K_{e_\t e_\t}$;
\begin{eqnarray}
&& K^{(PW,new)}_{\t\t}
= - K_{e_\t e_\t} \nonumber  \\
&&  =  -\frac{1}{2}\frac{(fg)^{1/2}}{  (f-g\  {{R_0}'(T)}^2)^{3/2}  } \times \nonumber \\
   &\times&   \Bigg\{
   \left(  \frac{{f,}_T}{f } - \frac{{g,}_T}{g}  \right) {R_0}' (T)  
   + \left( 2 \frac{{f,}_R}{f } - \frac{{g,}_R}{g}  \right) ({R_0}' (T))^2 \nonumber \\
   &+& \frac{{g,}_T}{f} {{R_0}'(T)}^3  -2 {R_0}''(T)
      \Bigg\} . 
\label{eq:K(PW)correct}
\end{eqnarray}
Comparing Eq.(\ref{eq:K(PW)correct}) with Eq.(\ref{eq:K(PW)}),  we realize that 
the term $-\frac{{f,}_R}{g}$ in  Eq.(\ref{eq:K(PW)})  should be removed and that the 
numerical factor ``2" in front of the factor $\frac{{g,}_T}{g}$ should be 
corrected to ``1". 

\subsection{Formulas for the pressures $p_z$ and $p_\phi$}

The expression for $\frac{dT}{dt}$ (corresponding to our $\frac{dt_+}{dt_-}$) 
in Ref.\cite{PeWa} has been corrected in Ref.\cite{PeWaE}, 
and the new expression coincides with our Eq.(\ref{eq:dTdt}). However, 
the expressions for $\frac{d^2T}{dt^2}$ and $r''_0(t)$ (corresponding to 
$\frac{d^2t_+}{dt_-^2}$ and $\ddot{\rho}_-(t_-) $, respectively)  
 in  Eq.(28) of Ref.\cite{PeWa} are not mentioned in Ref.\cite{PeWaE}; they actually should be 
corrected as our Eqs.(\ref{eq:dt2dt2}) and (\ref{eq:rhominus2dot}), respectively. 
The expression for $p_z$ has been corrected in Ref.\cite{PeWaE} as 
\begin{widetext}
\begin{equation}
\k p_z = \frac{e^{b(\xi_0)/2}}{\Delta (1-{{R_0}'}^2)^{3/2}}
\left\{
(1-\Delta) \left[ 
           \frac{\Delta}{R_0}(1-{{R_0}'}^2) - {R_0}''
             \right]
           -\frac{b'(\xi_0)}{2} (1- {R_0}')(1-{{R_0}'}^2)( {R_0}' - \Delta)   
 \right\} \ \ .
 \label{eq:PW-pz}
\end{equation}
\end{widetext}
However, it still needs modification (see our Eq.(\ref{eq:pz-result})), which is 
equivalent to replace the last two factors $(1-{{R_0}'}^2)( {R_0}' - \Delta)$ in Eq.(\ref{eq:PW-pz}) 
with $-{R_0}' [(\Delta +1) {R_0}' +1]$. 
The new expression of $p_z$ influences the dynamical equation for the shell, 
since the latter is obtained by setting $p_z=0$ 
(They only consider the case $p_z=0$). The rest is discussed in Sec.\ref{sec:4}.

The expressions  $p_\phi$ in  Eq.(29) of Ref.\cite{PeWa}
should also be modified as our Eq. (\ref{eq:pphi-result}).


\end{document}